\AtBeginDocument{%
    \addtolength{\footskip}{2.0\baselineskip}%
    \fancyfoot[L]{\textit{\textbf{Preprint --- do not distribute.}}}%
}
\documentclass[acmsmall,screen,authorversion,nonacm]{acmart}

\usepackage{csquotes}

\AtBeginDocument{%
  \providecommand\BibTeX{{%
    \normalfont B\kern-0.5em{\scshape i\kern-0.25em b}\kern-0.8em\TeX}}}

\setcopyright{acmcopyright}
\copyrightyear{2021}
\acmYear{2021}
\acmDOI{10.1145/1122445.1122456}

\acmJournal{JACM}
\acmVolume{37}
\acmNumber{4}
\acmArticle{6}
\acmMonth{10}

\usepackage[normalem]{ulem}
\usepackage{float}
\usepackage{array}
\usepackage{ctable} 
\usepackage{longtable}
\usepackage{multirow}

\useunder{\uline}{\ul}{}

\begin{document}




\title{Tasks, Time, and Tools: Quantifying Online Sensemaking Efforts Through a Survey-based Study}

\author{Andrew Kuznetsov}
\affiliation{
  \institution{Carnegie Mellon University}
  \city{Pittsburgh}
  \state{PA}
}
\email{kuz@cmu.edu}

\author{Michael Xieyang Liu}
\affiliation{
  \institution{Carnegie Mellon University}
  \city{Pittsburgh}
  \state{PA}
}
\email{xieyangl@cs.cmu.edu}

\author{Aniket Kittur}
\affiliation{
  \institution{Carnegie Mellon University}
  \city{Pittsburgh}
  \state{PA}
}
\email{nkittur@cmu.edu}

\renewcommand{\shortauthors}{Kuznetsov, et al.}


\begin{abstract}

Aiming to help people conduct online research tasks, much research has gone into tools for searching for, collecting, organizing, and synthesizing online information. However, outside of the lab, in-the-wild sensemaking sessions (with data on tasks, users, their tools and challenges) can ground us in the reality of such efforts and the state of tool support. We use a survey-based approach with aided recall focused on segmenting and contextualizing individual exploratory browsing sessions to conduct a mixed method analysis of everyday sensemaking sessions in the traditional desktop browser setting while preserving user privacy. We report data from our survey (n=111) collected in September, 2022, and use these results to update and deepen the rich literature on information seeking behavior and exploratory search, contributing new empirical insights into the time spent per week and distribution of that time across tasks, and the lack of externalization and tool-use despite widespread desire for support.

\end{abstract}



\maketitle

\section{Introduction}

People have many information needs that require them to search for and make sense of many web pages online, such as planning a personal trip or researching how to make a 3D game. Previous research has highlighted the importance of such exploratory search and sensemaking tasks in everyday web use, noting that they often involves more time, complex queries, processes, and goals \cite{marchionini_exploratory_2006} \cite{kellar2007field} and suggesting that such tasks are responsible for over a quarter of web searches \cite{donato_you_2010}. Furthermore, significant research has gone into tools for searching for, collecting, organizing, and synthesizing online information to help users with these tasks \cite{baldonado_sensemaker:_1997, dontcheva_summarizing_2006, chang_mesh_2020, kuznetsov_fuse_2022} (which we broadly term as part of the “sensemaking” process, as per \cite{pirolli_sensemaking_2005}).

However, several key questions remain open regarding how often people engage in sensemaking, how much time they spend on it, and what tools and processes they employ or wish to employ today. Many of the studies relevant to online sensemaking or exploratory searching were conducted more than a decade ago, while the internet and lives of those that use it have changed significantly, becoming increasingly fragmented while growing in size (241\% larger\footnote{Based on the long-running Netcraft Web Server Survey.} since the last study by Donato using Yahoo browsing history). It is possible that shortcuts such as knowledge cards and instant answers have changed the cost structure such that people just don’t often feel the need to "do the research themselves" even for open-ended tasks, and don’t spend much time on doing it when they do. The glut of information available today may be overwhelming, leading to more satisficing within sensemaking sessions as it becomes more difficult for people to even know if they have exhaustively covered the different perspectives and opinions and evidence, and which evidence they should believe - tracking all of which can quickly exceed working memory. Furthermore, it is unclear whether the prevalence of monetization through articles, affiliate advertising, and SEO for topics where people are looking for decision making support such as shopping or trip planning make things easier or harder for them. In this work we aim to investigate how people are handling the modern web and it's associated information overload; are they slowing down and spending more time; if and how are they externalizing information in order to track and synthesize it; and what are the opportunities here for tool support.

Answering these questions is challenging methodologically because of the tradeoffs involved in collecting data about not only what web pages users go to but their motivations, processes, mental models, and goals as they visit those pages. Web log analyses provide a comprehensive look at which pages users go to at scale, but require more indirect inference into what the user was actually thinking and doing. At the other end of the spectrum, interviews provide deep process insights but are high effort and often focus on a small sample of tasks (e.g., “tell us about the last time you did this”). Efforts to ground interviews in real usage, eg., asking participants to come in and “do a specific task” or “search as you normally would” may not be representative of their natural searching behavior and are limited in the amount of time researchers can see what they are doing. Conversely, diary studies can provide rich qualitative insights over a longer period of time but involve high effort in recruiting and coordinating users, and such sampling approaches may not provide a representative picture of any particular period of time. Finally, efforts involving researchers building specific tools for participants to use when browsing on their own (e.g., WebTracker \cite{choo1999information}, Vigo et al.'s Firefox add-on, \cite{vigo_real-time_2017}, Tabs.do, \cite{chang_tabs_2021}, Fuse \cite{kuznetsov_fuse_2022}) can be high cost to build and deploy as a data collection method, and users might not use them as they normally would their own browser.

In this paper, we aim to shed light on sensemaking efforts spent on the modern web through a survey-based approach with aided recall which focuses on segmenting and contextualizing individual exploratory web browsing sessions. This technique enables us to conduct a mixed method analysis of everyday sensemaking sessions in the traditional desktop browser setting while preserving user privacy. We use this methodology to report data from a survey of 111 participants about their online information seeking and sensemaking behavior, collected during the month of September, 2022. Our results regarding online sensemaking update and deepen the rich literature on information seeking behavior and exploratory search, contributing new empirical insights into the time spent per week and distribution of that time across tasks, as well as the lack of general tool use and externalization despite the widespread desire for tool support. As the online environment continues to rapidly evolve we hope our methodology and snapshot of behavior can be useful to researchers in providing ways to track web's evolution and understand changing user behavior with lower effort, while providing quantitative and qualitative insights into the contemporary web.

\section{Related Work}

\subsection{Characterizing Everyday Online Research}
To study how people navigate daily research tasks, prior research has leveraged various methods from in-person observation and retroactive log and survey analysis. In-person studies allow for convenient access to participants for follow-up questioning and enable researchers to control the research domain or task. Prior methods have included think-aloud, to examine how people's thought processes evolved throughout controlled lab studies or contextual inquiry sessions \cite{brandt_two_2009,capra_re-finding_2003,muramatsu_transparent_2001}, and to assess the utility of support tools \cite{palani2021conotate}. In some cases, researchers have invited participants into the lab setting to freely conduct searches without a designated set of topics \cite{byrne1999tangled,teevan_perfect_2004, kellar2007field}, a more natural way to capture open-ended user behavior compared to previous studies that used simulated work tasks \cite{borlund2003iir} or theory-inspired search tasks \cite{kelly2015development,o2020role} and pre-selected topics \cite{gadiraju2018analyzing,roy2021note}. However, such studies can present challenges in characterizing the extended browsing and searching processes that often span across multiple hours or days. Researchers have previously conducted in-person studies of information re-finding (e.g. \cite{capra2003re}, \cite{morris2008searchbar}) using multiple sessions scheduled a set number of days apart and studies of orienteering using unscheduled interruptions of co-located participants multiple times per day, but these techniques are unlikely to be feasible at scale. In Sellen et al.'s study of knowledge workers \cite{sellen_how_2002}, a two-day diary study, participants were interviewed at the end of the day in front of their workplace computer with their workday's browsing history open, although this would also be difficult to scale up to many users, and require additional privacy modifications for collecting non-workplace data. In person interviews can also involve significant costs in time and effort for the participants and researchers. Researchers looking to answer questions about web use with many participants have primarily utilized existing surveying efforts such as the semi-annual GVU WWW User Survey \cite{pitkow1994results} that ran in 10 editions from January 1994 to October 1998, using the datasets to speak to user strategies \cite{catledge1995characterizing}, web activities and goals, \cite{morrison_taxonomic_2001} and overall categories of web usage \cite{kehoe1996surveying}. More recently, researchers have used crowdsourcing platforms such as Amazon MTurk to survey participants about specific types of online tasks \cite{choi2023understanding}, although this specific work was limited to procedural 'how-to' knowledge and did not capture more than one task per participant. 

On the other hand, researchers have extensively investigated a variety of log data, such as people's browsing history \cite{matthijs_personalizing_2011}, tab usage \cite{chang_tabs_2021,chang_when_2021,huang_parallel_2010}, search queries \cite{donato_you_2010,spink2001searching,silverstein1999analysis,jansen2000real}, click-streams \cite{rozanski2001seize1,rozanski2001seize2}, client-side browser logs \cite{catledge_characterizing_1995, tauscher1997people,cockburn2001web} and interaction data \cite{guo_ready_2010} and outlined types of information that people commonly seek online and patterns in their search strategy and behavior. However, log data requires additional inference to identify people's personal intent and characteristics from usage telemetry. Researchers have previously explored augmenting log data collection by deploying pop-up surveys to search engine users randomly \cite{broder2002taxonomy}, but this method is rare and likely requires an established relationship between researchers and a search engine company to be feasible. Another recently popular method of collecting log data involves instrumenting the user's computer itself, such as configuring software trackers on work machines \cite{choo_information_2000}, or installing a specific 'tracked' browser \cite{zhang2020situ} or browser plug-in on a personal computer for use outside of the lab \cite{vigo_real-time_2017, van_kleek_finderskeepers_2011, kuznetsov_fuse_2022, jacucci2021entity}. Augmenting log data in this context has largely taken the form of follow-up interviews reviewing tracked data \cite{choo_information_2000,zhang2020situ} or prompting participants in real-time to answer questions when the tool detects a relevant activity \cite{vigo_real-time_2017}. However, such tool-based methods involve deploying customized applications to a diverse audience and require significant trust on the behalf of users as they give researchers direct access to their online activities. 

In this work, we devise a method for collecting online research data and records that combines the benefits of both log data and think-aloud studies through a survey-based study that can be quickly implemented and deployed by researchers. Our two-part survey asks users perform aided recall with their own personal web history and then elaborate on these tasks with closed and open-ended questions, enabling a mixed method analysis of in-the-wild sensemaking sessions at a larger scale than typical in-person tasks and with more user insights than log data.

\subsection{Sensemaking in Online Research Tasks}

In this work, we aim to paint a more recent picture of the online research tasks and sensemaking time of daily internet users with finer granularity than previous work. We use the term sensemaking as defined in broad terms, as the developing of a mental model of an information space in service of a user's goals, which can include the activities of information seeking, exploratory search, collecting and organizing information, as well as synthesis activities \cite{pirolli_sensemaking_2005,dervin_overview_1983,marchionini_information_1995,hahn_knowledge_2016}. Prior research  suggested that the users conducting online research typically 'loop' back and forth \cite{russell_cost_1993} between a \textit{information seeking} phase, in which people search and collect information from various sources, and a focused \textit{sensemaking} phase, in which people synthesize the collected information into mental structures and schema \cite{pirolli1999information}. Studies from more than a decade ago have indicated that people tend to spend the majority of time seeking new information \cite{pirolli_information_1999,marchionini_information_1995,cepeda_distributed_2006}, whether it is for everyday decision making \cite{payne1991consumer,head2011college,savolainen2005everyday}, or tasks in specific domains like researching medical diagnoses \cite{billman_medical_2007} or solving programming problems \cite{brandt_two_2009}. However, the web has since gone through tremendous changes and become all-encompassing with rich information and complex applications, and people's web usage behavior and browsing patterns have also seen massive shifts \cite{chang_tabs_2021,huang_parallel_2010,chang_when_2021} from navigating web directories \cite{noauthor_dmoz_nodate} and viewing individual web pages to exploring many resources in parallel for complex sensemaking tasks both on desktop and mobile\cite{hahn_bento_2018,liu_reuse_2021,chang_searchlens_2019}.

Previous work has also theorized the higher-level processes (such as needs and goals) that drive user actions in online research tasks, largely converging on the central idea that people's needs are dynamic and continuously evolve \cite{teevan_perfect_2004}. For example, Wilson and Walsh \cite{wilson_information_1997,wilson_user_1981} stressed the influence of the underlying context and need on search behavior, while O'Day and Jeffries \cite{oday_orienteering_1993} summarized common triggers and stop conditions that propel people's information seeking progress. Agarwal et al. \cite{agarwal_making_2012} further modeled the searching and making sense of information as a way to address the gaps and discontinuities in one's knowledge. Our research deepens and enriches this work by providing mixed-method quantitative and qualitative insights into users' online sensemaking needs that is fine-grained, comprehensive, and reflective of a more modern web information environment.

\subsection{Augmenting Online Sensemaking}

Many exploratory search and information-seeking tools exist that accelerate the online research process by supporting activities such as seeking out new information, exploring relevant domains, and filtering data for integration. For scope, in this section we focus on tools that support the sensemaking phase most directly by helping users ingest results, ruminate on their impact, and digest them to provide information for the task at hand. Tools that help people better capture information online as they see it, such as Hunter Gather \cite{schraefel_hunter_2002}, SenseMaker \cite{baldonado_sensemaker:_1997}, Clipper \cite{kittur_costs_2013}, Wigglite \cite{liu_wigglite_2022}, and Pocket enable users to voluntarily keep track of bits and pieces of information or entire webpages and documents for later consumption. To improve the information collection efficiency and keep people in the flow, prior work has introduced semi-automatic systems such as Dontcheva et al.'s web summarization tool \cite{dontcheva_summarizing_2006} and Thresher \cite{hogue_thresher_2005}, which let users create and curate patterns and templates of information that they would like the system to automatically collect as they browse other webpages. Prior work has also introduced various ways to support in-depth organizing and structuring of results once collected. For example, Adamite \cite{horvath_understanding_2022} encourages the user to categorize an information clip immediately after capture. Unakite \cite{liu_unakite:_2019} and Mesh \cite{chang_mesh_2020} enable users to build comparison tables of various options and criteria. ForSense \cite{rachatasumrit_forsense_2021} leverages natural language processing to automatically cluster information clips based on themes and topics while Fuse \cite{kuznetsov_fuse_2022} enables users to create deeply nested hierarchies similar to a file system. 

However, the widespread impact of online sensemaking tools in the wild has largely remained limited, with some exceptions including general use notetaking tools \cite{van_kleek_finderskeepers_2011} and clipping tools \cite{kuznetsov_fuse_2022} examined in deployment studies for longer periods of time. In addition, prior research on the usage of browser history (a passive device to help people recall and revisit) and bookmarks (a default browser feature to help with information collection) have found little usage or high abandonment after initial use \cite{weinreich_off_2006,aula_information_2005,jones_once_2002,jones_keeping_2001,byrne_tangled_1999} while simultaneously finding little evidence of customized tools beyond what users have at hand \cite{capra2010tools}. Our study aims to further investigate the gap of complex sensemaking tools in general usage, and  shed light on people's real-world tool choices, usage patterns, wants and needs, in order to surface design implications for next generation of tool support.


\section{Sensemaking Survey}

In order to collect accurate information regarding user sensemaking time and tasks, we designed a two-part survey with the ability to cross reference results between the sections. In doing so, we also needed to balance between accuracy and privacy considerations for users, so that users felt comfortable sharing their personal browser activities and history logs. We discuss this composite design which combines both user-reported and log data in two parts: a high-level description of their recent tasks and second deep dive into individual tasks that involves copying-and-pasting their related browser history.

\subsection{Survey Design}
In the first section, we aimed to understand the full set of activities users engaged with online and how much time they spent on each. To decrease user effort, we ask that users group their online search history into 'online research tasks' from the past 72 hours and estimate the time spent on each into 1/5/15/30 minute increments. To ensure that users were accurately completing this task, we provided example online research tasks such as 'researched a camera' and 'planning a vacation/purchase' and examples of tasks that were not relevant such as 'copyediting a google doc or browsing social media'. Asking for high-level user estimates of time in this way eliminated the need for us to engage in the difficult task of computationally segmenting users' browsing history while keeping the task relatively light for survey participants. Early versions of the survey conducted as in-person interview study demonstrated that asking users to record online research tasks using the last 30 days of browser history took about 30 minutes in practice, which we deemed infeasible for a larger survey. Another subsequent version of the survey, this time conducted through Prolific, found a shortened period of two weeks to be similarly infeasible for large-scale collection. Although this much longer data is not reported in this study, it became invaluable as a baseline for determining our final collection period for the study. A final collection window of seventy-two hours was chosen by cross-referencing time period data from the first and second survey sections. To do this, we compared how far back participants reported five appropriate tasks in the first section with the average time period over which participants reported their 'five most recent online research tasks' in the second section, finding that 72-hours would be long enough to provide overlap between the two sections. Additionally, collecting 72 hours allows us to ensure that at least one of the days of the collection window does not land on the weekend, when user behavior may be different. Prior literature analyzing logs of search engine users \cite{donato_you_2010} and study participants \cite{ghosh2018searching} also used similar time periods, suggesting to us that this scope would be sufficient to capture longer tasks. In total, our prompt to users read as follows: "For the past three (3) days, please record each online research task with a brief description (4-6 words) and the estimated time you spent on it. For example, “Searched for a new camera, 1 hour”.  Don’t worry about estimating precisely, round off to the nearest hour or if less than an hour whether it took 1/5/15/30 minutes." To capture additional contextual information, we also included the instructions: "If the task is part of a longer task, please include your best estimate of the longer task as well, e.g., “Searched for a new camera, 1 hour, part of a larger task, 6 hours”. As previously mentioned, we also provided users the option of 'censoring a task' by writing 'private' for its description but asked that users to fill out the time estimate. This enabled us to give participants the option of giving us a near-complete picture of their activities while giving participants the option of maintaining privacy. This was followed by an example input with one example of a normal task, a task part of a larger task, and a 'private' task. There were several risks with this approach, one of which is that users may claim their entire set as private to hide a low-effort response. However, in practice users chose to omit less than <1\% of user tasks. Additionally, we were concerned that the estimates users would give would be inaccurate; to test this in practice, a random sample of collected survey data was cross-referenced between the first 'high-level' section and the second 'in-depth' section. Results from the first '72-hour' section of the survey were reviewed and matched to the best of ability with the copy-pasted history logs of the 'most recent tasks' from the second section, ultimately finding that users were accurate in their assessments. Although reassuring for our methodology, this was largely unsurprising since participants in the first section were asked to open and reference their browser history data while reporting their estimates. 

\begin{figure*}[h]
\centering
\includegraphics[width=\textwidth]{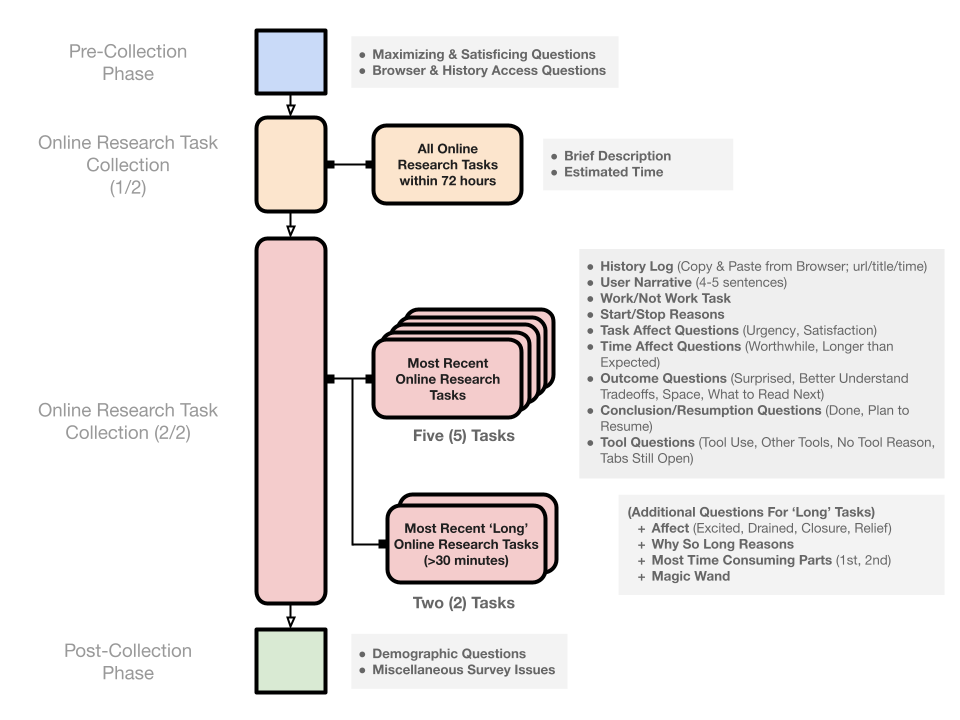}
\vspace{-5mm}
\caption{An overview of this study's survey method: a pre-collection phase, followed by a two-part collection of online research tasks, followed by a post-collection phase. A high level overview of the questions in each of the phases is enumerated in the grey boxes to the right of each.  
}
\end{figure*}

In the second section, we aimed to deeply dive into the context of individual tasks using log data as a scaffold for users, and a source of validation after it was collected. In order to ensure that we collected a diversity of tasks, we asked users to consider their five most recent tasks, as well as two additional tasks that were over 30 minutes for a total of seven distinct tasks. For each of these seven, participants were asked to first copy-and-paste in the relevant browser history entries (including the website's \texttt{url}, title , and timestamp of visit) from their browser's web history page, and then manually add additional context by responding to the prompt: "imagine you are helping someone to repeat the same task. Write a list of steps (at least 4-5 sentences) that would help them do it as closely as possible to the way you did it.". This last section would serve as an additional point of reference in understanding the task, the users goals and intermediate actions. Afterwards, participants were asked a series of questions to reflect about the task in detail, such as why they started and paused/stopped (if applicable) the task, which part they thought took the most amount of time or effort, and what section of the task they would want solved in the hypothetical situation that they were given a magic wand, etc. Similar to the previous survey section, participants were allowed to omit tasks - this was done by recording an omission and substituting the task with the next available task chronologically. In practice this was rarely done, with participants each omitting 0.18 tasks on average, which when accounting for the amount of tasks collected means participants omitted one task per 38 tasks reported. To double check that this process of chronological substitution did not result in users reporting very old tasks they may no longer have the context for, we also asked participants to report the date at which the task had taken place. In practice, we found that all tasks used in the second 'copy and paste' phase were within the past 30 days. Long tasks were also verified as being over 30 minutes by inspecting timestamps within the collected browser history logs. The survey then wrapped up with eliciting participant's demographic information - the full protocol can be found in the Appendix. The survey was deployed via Google Forms. Each survey took participants around an hour to finish, and they were compensated \$15 per hour for their time.

\subsection{Participants}
To gain a diverse sample of sensemaking users and tasks, we conducted a survey with 139 participants recruited from an online survey pool using Prolific (Ages: 18-60+; 18 - 24, 18.0\%; 25 - 39, 60\%; 40 - 59, 20\%; 60 +, 2.7\%; 41\% female; recruited from US-based participants fluent in English) over the course of 10 days. In line with prior work, participants self-reported as coming from a wide variety of backgrounds, the most popular being self-employed (11\%), student (10\%), manager (9\%), retail (6\%), and engineering (5\%) of the 74 distinct positions reported which included a paralegal, delivery driver, and teaching assistant professor. Furthermore, 32\% had a high school degree or equivalent, 44\% of the participants had a Bachelor's degree, and 8\% had a master's degree or higher graduate degree. We believe that this diversity in participants' backgrounds helped our survey capture a variety of user profiles and online sensemaking behaviors.  

\subsection{Validity Checks}
As previously mentioned, the dataset was rigorously reviewed to ensure that participants had correctly followed instructions and uploaded seven distinct 'online research' tasks, with the last two being longer than 30 minutes. This was done by reviewing the time-stamps provided by the search history data. Additionally, researchers reviewed all submitted tasks in the first and second phase to verify that submitted tasks were indeed 'online research tasks' that would fit the description of tasks involving 'sensemaking' as per Russell et al. \cite{russell_cost_1993}. This was done by reviewing the survey responses between two members of the research team and excluding low-quality or simple 'retrieval' tasks such as looking up the date of a football game. All non-relevant tasks in the first section were excluded from calculations regarding time, topic, and quantity of sensemaking tasks. The presence of such tasks in the second section resulted in the exclusion of the entire participant in any calculations. We found that the small amount of participants who did not report seven distinct tasks that fit the requested description were entirely comprised of 'low data quality' participants, largely from misinterpreting the instructions, attempting to provide the same task seven times, a technical issue in retrieving search history, or providing low effort responses to follow-up questions. To isolate any technical and logistical issues in accessing browser history that may affect data validity, participants were asked to report any issues in accessing the browser history from any computers or browsers they use on a daily basis. Although still paid for their time, these participants were excluded in calculations to better estimate everyday sensemaking efforts. After these data validity checks, the final dataset contained 111 participants.

\section{Survey Results}

\subsection{How much time do people spend on sensemaking efforts on the web?}


While large scale log data analyses have shown that sensemaking-related searches can comprise over quarter of web searches \cite{donato_you_2010}, it is hard to ascertain how much of their actual time users spend on sensemaking tasks without direct feedback from them, as certain tasks may be lumped together in their minds, they may take offline breaks their browser isn't aware of, or for other reasons not visible in the data.  If these searches are components of larger sensemaking tasks, it is possible their prevalence may be being overestimated; conversely, if users spend more time reading and digesting online information for these sensemaking tasks than simple information seeking tasks they may be undercounted in terms of their importance in users' daily behavior. Our study aims to estimate the number of tasks and amount of time spent on those tasks defined by users' own mental models of their tasks and goals and grounded in a comprehensive (but time-delimited) interval of their own browsing history, as opposed to approaches such as log analysis, interviews, or sampling or diary studies that do not provide this combination of data characteristics.

\begin{figure*}[h]
\centering
\includegraphics[width=\textwidth]{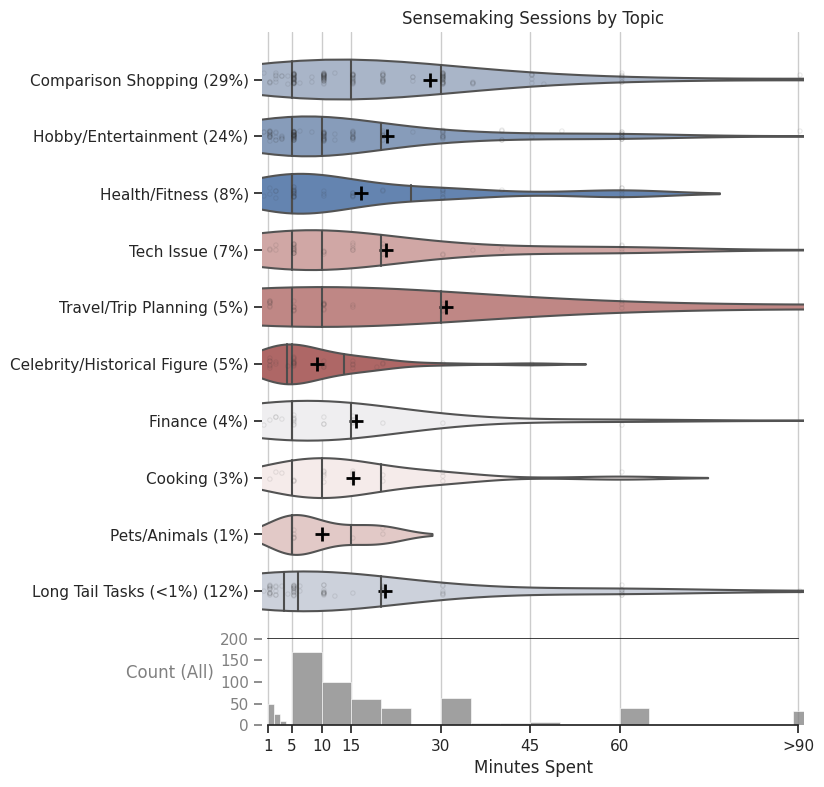}
\vspace{-5mm}
\caption{A violin plot of the sensemaking sessions by topic, sorted from top to bottom by average number of tasks in the dataset. Vertical gray bars within each plot indicate quartiles, a '+' symbol within each plot represents the average session duration. Individual sessions are plotted using a strip plot with transparent circles designating each session. Task categories below 1\% are represented at the bottom in the 'Long Tail Tasks' plot. Further below, a histogram of session durations guides the reader in understanding the distribution of the data.
}
\end{figure*}

Overall, participants reported spending an average of 4.4 hours a week conducting an average of 12.5 exploratory web tasks weekly. This data, calculated from 111 in-the-wild participants over a 72-hour period, corresponds to an average of about 230 hours and 651 tasks a year. Most of the collected explorations were much longer than a typical website session at 19 minutes, bringing them in line with industry estimates of the average Google search session at 21.5 minutes according market research published by Statista\footnote{Based on "Most popular websites worldwide as of November 2022, by time per visit" from https://www.statista.com/statistics/1201901/most-visited-websites-worldwide-time-visit/}. Within the collected tasks, a majority of participants' time was spent on a relatively small proportion of long tasks, suggesting we had captured long-running exploratory web browsing behavior in our survey. These longer (>=30 minute) tasks, despite comprising only 23\% of surveyed tasks, consumed 72\% of surveyed time. Furthermore, a minority of the surveyed tasks were terminated quickly - 35\% were completed in 5 minutes or less, and 13\% were completed nearly instantly in two minutes or less. Our survey enabled us to examine if these short research tasks may have originated by reporting fragments of a larger task, however only 2\% of the short (<30 minute) tasks were reported by participants to be part of a larger (>=30 minute), suggesting the presence of distinct short-running research tasks and minimal fragmentation in reporting longer tasks.

\subsection{Overview of Task Types}

Additionally, the authors coded and assigned task types that encode both domain and user intentions, utilizing three passes of coding (one open, one closed, one conciliatory) between two members of the research team. In the first pass, researchers performed open-coding, generating a list of task categories. After this, researchers collaborated to create a shared set of closed codes. In the second pass, researchers reviewed task descriptions and history logs to assign a closed code to each reported task. After this, in the third conciliatory pass, researchers discussed any discrepancies and agreed on an assigned code. The number of deviations before conciliation between the closed codes was minimal (<10 tasks overall, Cohen's Kappa $\sim$99\%), resulting in a short discussion time and full agreement following the pass. The resulting task types and their frequency are noted here: Comparison Shopping (29\%), Researching a hobby/entertainment (24\%), Researching health/fitness (8\%), Diagnosing a technology issue (7\%), Trip Planning (5\%), Learning about a celebrity/historical Figure (5\%), Finance-related (4\%), Cooking-related (3\%), and Pet-related (1\%). Domains under <1\% are not reported. We also report the time spent on the above tasks as a percentage of total time spent: Comparison Shopping (38\%), Researching a hobby/entertainment (23\%), Researching health/fitness (7\%), Diagnosing a technology issue (6\%), Trip Planning (8\%), Learning about a celebrity/historical Figure (2\%), Finance-related (3\%), Cooking-related (2\%), and Pet-related (1\%). We note that a majority of user time captured is in the personal domains of comparison shopping and researching a hobby and entertainment, as it's likely that participants were not reporting work-related tasks in this survey. Unless explicitly mentioned, the statistics reported past this section related to those computed from the 5 short and 2 long tasks that involved copying-and-pasting history logs and answering follow-up questions.

\subsection{Forces that sustain sensemaking efforts}

To understand what keeps people engaged in sensemaking efforts, we solicited the reasons why participants felt their tasks took so much time. We find that participants felt compelled to continue searching for information with the following concerns: doing a thorough job (42\%), potential to find something better (25\%), missing out on something (23\% ), not finding something they were expecting to find (17\%) and being unable to find something that matches their preferences yet (15\%). However, participants also reported several motivating forces that kept them searching, for example, finding new results was rewarding (28\%), and exploring the topic was interesting to them (2\%). Lastly, other reasons for participants to keep searching include the lack of an explicit stopping point (23\%), the nature of the task was time consuming (14\%), and a drive for further clarity (<1\%). 

Our results suggest that people feel an obligation to explore a never-ending stream of results until they have reached the perception of saturation. This saturation point is driven chiefly by several forces, such as the need to reassure oneself that they have done due diligence, the excitement of new results, and the fear of missing a good, better, or previously-expected result, and is achieved largely by a perception of decreased discovery of new, quality results, and further laid to rest by running out of energy to continue or being interrupted by chores or distractions. This is further elaborated in Tables 1 and 2. Our survey enabled participants to specify multiple concerns per task, as a result the following statistics add up to more than 1. In all, it is clear that the need to continue searching for information is largely driven by the perception of obligation - the need to do a thorough job, with maximization behaviors  (better, missing out on something) and fun takes a close second seat. This has further implications regarding the importance of how people perceive a specific domain (such as health) and the information found in that domain (such as regulated health warnings), and how it may uniquely shape sensemaking in that domain. We discuss this in detail in section \ref{sec:discussion}. 

\subsection{Forces that deescalate sensemaking efforts}

On the other hand, our survey also uncovered the reasons why people decided to stop searching for information. Out of the 777 responses we gathered, ``seeing most of the good results'' (45\%) dominated the reasons for terminating a sensemaking task. Conversely, unexpected distraction such as self-initiated breaks (14\%) and routine tasks (13\%), in-person and digital distractions (11\%, 9\%) were rarer, on par with re-prioritization or task switching (12\%), exhaustion (11\%) and task completion (9\%). Being blocked by waiting for someone's feedback (5\%) or feedback from technology (1\%) was relatively rare in our dataset. Specifically in longer tasks, exhaustion (19\%) and real world distraction (14\%) were more common, but seeing most of the good results still remained the primary reasons user stopped their tasks (37\%). These results suggest that internal factors (e.g. running out of good answers or energy) rather than external influences such as distractions, or re-prioritization dominate the decision to stop a sensemaking activity. Further implications are elaborated in the discussion. 

We conducted an additional analysis regarding user task completion as the previous results note that participants explicitly stop because they feel finished (9\%) almost as often as running out of energy and quitting (10\%). At first glance, this suggests that very large majority (around 90\%) of tasks are left unfinished. Our survey had an additional point of reference for understanding the termination of tasks.  Participants were asked to rate on a 7-point likert scale to rate if they felt done with the task, satisfied with what they had found so far, and whether they were planning to resume the task. Overall, a far majority of participants reported feeling satisfied to some degree (86\% average) after finishing their session. Observing the scores for ``being done'' (strongly agree: 49\%, moderately agree: 14\%, slightly agree: 8\%, neutral: 8\%), we see that in more than 60\% of cases, participants feel fairly strongly that they are done with their task, suggesting that a user's explicit decision to prematurely terminate a task because of full satisfaction plays a fairly small part in the completion of a task, and, instead, participants situationally wind down tasks and consider them complete, with some degree of satisfaction. This is further supported with the moderate amount of participants planning to resume or revisit the task at a later date (42\%) in some form. Overall, our data suggests that sensemaking efforts are rarely explicitly stopped due to completion, but rather left in a state of satisfaction and potentially revisited at a future date. Additional likert scale questions present in the long tasks further confirm this, with only about half of tasks (52\%) leaving participants with some amount of closure.

\subsection{Tool Usage}
\begin{table}[]
\begin{tabular}{|lr|}
\hline
\multicolumn{2}{|l|}{\textbf{Why do you feel like you ended up spending so much time? }} \\ \hline
\multicolumn{1}{|l|}{Felt like I hadn’t done a thorough job yet}                      & 41\% \\ \hline
\multicolumn{1}{|l|}{Finding new results/options was rewarding}                         & 28\% \\ \hline
\multicolumn{1}{|l|}{Felt like I could find something better than what I first found}        & 25\%       \\ \hline
\multicolumn{1}{|l|}{Felt like I was missing out on something}                        & 23\% \\ \hline
\multicolumn{1}{|l|}{Did not have a explicit stopping point}                          & 22\%          \\ \hline
\multicolumn{1}{|l|}{Felt like I hadn’t found something I was expecting to find yet}  & 17\%          \\ \hline
\multicolumn{1}{|l|}{Everything I was finding did not match my constraints/preferences} & 15\%          \\ \hline
\multicolumn{1}{|l|}{Did not have a concrete goal}                                    & 15\%          \\ \hline
\multicolumn{1}{|l|}{The content/topic was interesting}                                 & 2\%          \\ \hline
\multicolumn{1}{|l|}{The nature of the research was just very time consuming}           & 1\%          \\ \hline
\multicolumn{1}{|l|}{Needed a little {[}more{]} clarity}                              & <1\%          \\ \hline
\end{tabular}
\vspace*{3mm}
\caption{User responses for continued engagement in sensemaking efforts, collected from 222 'long task' responses from users.}
\end{table}

\begin{table}[]
\begin{tabular}{|lrrr|}
\hline
\multicolumn{4}{|c|}{\textbf{Why did you stop working on this task? (by task type)}} \\ \hline
\multicolumn{1}{|l|}{} &
  \multicolumn{1}{r|}{All} &
  \multicolumn{1}{r|}{Short (N=555)} &
  \multicolumn{1}{r|}{Long (N=222)} \\ \hline
\multicolumn{1}{|l|}{Seemed to have seen most of the good results} &
  \multicolumn{1}{r|}{\textbf{45\%}} &
  \multicolumn{1}{r|}{\textbf{49\%}} &
  \multicolumn{1}{r|}{\textbf{36\%}} \\ \hline
\multicolumn{1}{|l|}{Stopped to take a break (e.g. to browse social media)} &
  \multicolumn{1}{r|}{\textbf{14\%}} &
  \multicolumn{1}{r|}{\textbf{14\%}} &
  \multicolumn{1}{r|}{12\%} \\ \hline
\multicolumn{1}{|l|}{Stopped to do routine task (e.g. checking email)} &
  \multicolumn{1}{r|}{\textbf{13\%}} &
  \multicolumn{1}{r|}{\textbf{14\%}} &
  \multicolumn{1}{r|}{11\%} \\ \hline
\multicolumn{1}{|l|}{Something else became higher priority} &
  \multicolumn{1}{r|}{12\%} &
  \multicolumn{1}{r|}{12\%} &
  \multicolumn{1}{r|}{12\%} \\ \hline
\multicolumn{1}{|l|}{Real world distraction (e.g. someone stopping by)} &
  \multicolumn{1}{r|}{11\%} &
  \multicolumn{1}{r|}{10\%} &
  \multicolumn{1}{r|}{\textbf{14\%}} \\ \hline
\multicolumn{1}{|l|}{Ran out of energy to continue} &
  \multicolumn{1}{r|}{11\%} &
  \multicolumn{1}{r|}{7\%} &
  \multicolumn{1}{r|}{\textbf{19\%}} \\ \hline
\multicolumn{1}{|l|}{Other: I was done / found what I was looking for} &
  \multicolumn{1}{r|}{9\%} &
  \multicolumn{1}{r|}{9\%} &
  \multicolumn{1}{r|}{9\%} \\ \hline
\multicolumn{1}{|l|}{Online distraction (e.g. a notification)} &
  \multicolumn{1}{r|}{9\%} &
  \multicolumn{1}{r|}{9\%} &
  \multicolumn{1}{r|}{8\%} \\ \hline
\multicolumn{1}{|l|}{Stopped to wait for someone else} &
  \multicolumn{1}{r|}{5\%} &
  \multicolumn{1}{r|}{4\%} &
  \multicolumn{1}{r|}{6\%} \\ \hline
\multicolumn{1}{|l|}{Stopped to wait for a website/computer/program} &
  \multicolumn{1}{r|}{1\%} &
  \multicolumn{1}{r|}{1\%} &
  \multicolumn{1}{r|}{1\%} \\ \hline
\end{tabular}
\vspace*{3mm}
\caption{User responses for reasons terminating sensemaking efforts per task. The top three reasons are bolded to assist in comparison.}
\end{table}
Our results regarding tool use provide preliminary evidence that most online sensemaking tasks are completed without the usage of tools. As shown in table 3, by far the most common response was to not use tools and attempt to keep everything in their heads (76\%, high even when only looking at long tasks: 69\%).  The most popular tools overall were documents (10\%), physical paper (8\%) and spreadsheets (4\%). A long tail of tools below 1\% of reported use follows this, from various note-taking capabilities of the phone (0.6\%) and browser functionality such as the clipboard (0.1\%), bookmarks (0.1\%), and tabs (0.3\%), to one-off responses relating to Youtube playlists, calculators, and screenshots. It is worth noting that this study observed participants in a desktop setting, and may not generalize to a mobile setting.

The low amount of tool use, especially in long tasks, could be seen as surprising, as participants noted that these tasks make them feel drained to some degree after their search (38\%). We followed up on this in our survey, asking participants why they chose not to use any tools if they answered ``kept everything in my head'' for the tool use question. The results, viewable in table 4, contains the many reasons why participants decided not to use any tools as they completed their task. A few reasons dominate: confidence in recalling everything (34\%), the lack of a perceived need to save any answers (33\%), followed by the lack of good answers to save (13\%) or energy to put the information into the tool (14\%). Focusing on the task at hand (11\%), and a feeling of urgency (8\%) were next, followed by high activation energy (3\%), feeling overwhelmed (2\%), no obvious place to store things (2\%), the format of the information (1\%) and cluttering an existing tool (1\%). These results suggest that the primary reason people do not use support tools is that they don't believe they need them, despite their evident challenges with keeping everything in their heads.

\begin{table}[]
\begin{tabular}{|lrrr|}
\hline
\multicolumn{4}{|c|}{\textbf{What other software, if any, did you use to help manage your task? (by task type)}} \\ \hline
\multicolumn{1}{|l|}{} &
  \multicolumn{1}{r|}{All} &
  \multicolumn{1}{r|}{Short (N=555)} &
  \multicolumn{1}{r|}{Long (N=222)} \\ \hline
\multicolumn{1}{|l|}{I kept everything in my head} &
  \multicolumn{1}{r|}{\textbf{76\%}} &
  \multicolumn{1}{r|}{\textbf{79\%}} &
  \textbf{69\%} \\ \hline
\multicolumn{1}{|l|}{A document (e.g., a Google or Word doc)} &
  \multicolumn{1}{r|}{\textbf{10\%}} &
  \multicolumn{1}{r|}{\textbf{9\%}} &
  \textbf{13\%} \\ \hline
\multicolumn{1}{|l|}{A piece of paper} &
  \multicolumn{1}{r|}{\textbf{8\%}} &
  \multicolumn{1}{r|}{\textbf{7\%}} &
  \textbf{11\%} \\ \hline
\multicolumn{1}{|l|}{A spreadsheet}       & \multicolumn{1}{r|}{4\%}  & \multicolumn{1}{r|}{4\%}  & 5\%  \\ \hline
\multicolumn{1}{|l|}{Other software/website (e.g. Notion)} &
  \multicolumn{1}{r|}{2\%} &
  \multicolumn{1}{r|}{2\%} &
  3\% \\ \hline
\multicolumn{1}{|l|}{Copied and pasted}   & \multicolumn{1}{r|}{1\%}  & \multicolumn{1}{r|}{1\%}   & 1\%  \\ \hline
\multicolumn{1}{|l|}{Browser bookmarks}   & \multicolumn{1}{r|}{1\%}  & \multicolumn{1}{r|}{1\%}   & 2\%   \\ \hline
\multicolumn{1}{|l|}{Phone}               & \multicolumn{1}{r|}{1\%}  & \multicolumn{1}{r|}{1\%}  & <1\%  \\ \hline
\multicolumn{1}{|l|}{YouTube playlist}    & \multicolumn{1}{r|}{<1\%}  & \multicolumn{1}{r|}{<1\%}  & -       \\ \hline
\multicolumn{1}{|l|}{Physical planner}    & \multicolumn{1}{r|}{<1\%}  & \multicolumn{1}{r|}{<1\%}  & -       \\ \hline
\multicolumn{1}{|l|}{Amazon favorites}    & \multicolumn{1}{r|}{<1\%}  & \multicolumn{1}{r|}{<1\%}  & -       \\ \hline
\multicolumn{1}{|l|}{Saved coupon codes}  & \multicolumn{1}{r|}{<1\%}  & \multicolumn{1}{r|}{<1\%}  & -       \\ \hline
\multicolumn{1}{|l|}{Keeping tabs open}   & \multicolumn{1}{r|}{<1\%}  & \multicolumn{1}{r|}{<1\%}  & <1\%  \\ \hline
\multicolumn{1}{|l|}{Downloads}           & \multicolumn{1}{r|}{<1\%}  & \multicolumn{1}{r|}{<1\%}  & -       \\ \hline
\multicolumn{1}{|l|}{Screenshots}         & \multicolumn{1}{r|}{<1\%}  & \multicolumn{1}{r|}{<1\%}  & -       \\ \hline
\multicolumn{1}{|l|}{Facebook messenger}  & \multicolumn{1}{r|}{<1\%}  & \multicolumn{1}{r|}{<1\%}  & -       \\ \hline
\multicolumn{1}{|l|}{Canvas LMS}          & \multicolumn{1}{r|}{<1\%}  & \multicolumn{1}{r|}{<1\%}  & -       \\ \hline
\multicolumn{1}{|l|}{Outlook tasks}       & \multicolumn{1}{r|}{<1\%}  & \multicolumn{1}{r|}{-}       & <1\%  \\ \hline
\multicolumn{1}{|l|}{Calculator}          & \multicolumn{1}{r|}{<1\%}  & \multicolumn{1}{r|}{-}       & <1\%  \\ \hline
\multicolumn{1}{|l|}{Browser autofill}    & \multicolumn{1}{r|}{<1\%}  & \multicolumn{1}{r|}{<1\%}  & -       \\ \hline
\end{tabular}
\vspace*{3mm}
\caption{User responses for tools used. The top three reasons are bolded to assist in comparison.}
\end{table}

\begin{table}[]
\begin{tabular}{|lrrr|}
\hline
\multicolumn{4}{|c|}{\textbf{If you didn’t use anything, what prevented you from doing so? (by task type)}} \\ \hline
\multicolumn{1}{|l|}{} &
  \multicolumn{1}{r|}{All} &
  \multicolumn{1}{r|}{Short} &
  \multicolumn{1}{r|}{Long} \\ \hline
\multicolumn{1}{|l|}{Confident able to recall it without writing it down} &
  \multicolumn{1}{r|}{\textbf{34\%}} &
  \multicolumn{1}{r|}{\textbf{38\%}} &
  \textbf{27\%} \\ \hline
\multicolumn{1}{|l|}{Didn’t feel the need to save the answer/information anywhere} &
  \multicolumn{1}{r|}{\textbf{33\%}} &
  \multicolumn{1}{r|}{\textbf{35\%}} &
  \textbf{27\%} \\ \hline
\multicolumn{1}{|l|}{Didn’t feel that found anything useful/worth keeping} &
  \multicolumn{1}{r|}{\textbf{13\%}} &
  \multicolumn{1}{r|}{\textbf{14\%}} &
  \textbf{11\%} \\ \hline
\multicolumn{1}{|l|}{Didn’t believe it warranted effort to put info. in another tool} &
  \multicolumn{1}{r|}{12\%} &
  \multicolumn{1}{r|}{14\%} &
  9\% \\ \hline
\multicolumn{1}{|l|}{Was an urgent task} &
  \multicolumn{1}{r|}{8\%} &
  \multicolumn{1}{r|}{8\%} &
  10\% \\ \hline
\multicolumn{1}{|l|}{Wanted to focus on completing the task} &
  \multicolumn{1}{r|}{11\%} &
  \multicolumn{1}{r|}{11\%} &
  \textbf{11\%} \\ \hline
\multicolumn{1}{|l|}{Would have required more energy than had at the moment} &
  \multicolumn{1}{r|}{3\%} &
  \multicolumn{1}{r|}{3\%} &
  3\% \\ \hline
\multicolumn{1}{|l|}{Feeling too overwhelmed to do it} &
  \multicolumn{1}{r|}{2\%} &
  \multicolumn{1}{r|}{1\%} &
  4\% \\ \hline
\multicolumn{1}{|l|}{Felt that there was no obvious place to save/store things} &
  \multicolumn{1}{r|}{2\%} &
  \multicolumn{1}{r|}{3\%} &
  2\% \\ \hline
\multicolumn{1}{|l|}{The type/format of the answer/info made it hard to record} &
  \multicolumn{1}{r|}{1\%} &
  \multicolumn{1}{r|}{1\%} &
  1\% \\ \hline
\multicolumn{1}{|l|}{Worried about cluttering up an existing tool I use} &
  \multicolumn{1}{r|}{1\%} &
  \multicolumn{1}{r|}{1\%} &
  1\% \\ \hline
\end{tabular}
\vspace*{3mm}
\caption{User responses for a lack of tool use during sensemaking efforts. The top three reasons are bolded to assist in comparison.}
\end{table}

\subsection{Tool Needs and Wants}

To further understand what types of features would have been perceived useful to respondents, participants were asked to elaborate on the hardest and second hardest part of the task after each of the ``long'' tasks. We asked for a second hardest result to get participants to think more critically by prompting them to review its relation to another difficult task. Additionally, participants were asked to describe a segment of the task that they could solve using a magic wand, a typical user experience needs gathering prompt. Similar to previous coding in this work, the authors coded the aggregated results using two passes of coding (one open, one closed) with discussion to enable reconciliation. This yielded 12 themes for needs and 16 themes for use wants from the 222 qualitative results coded. We report these below:

Participants mentioned various initial parts of the foraging process as difficult: \textit{consuming} (e.g. reading, watching videos, consuming in different formats, reading reviews, managing a large amount of results), \textit{obtaining sufficient breadth} (e.g. saturating info to a full understanding, capturing important symptoms, looking at all the different options, getting an understanding of the space, diving too deeply on first results found), and \textit{ensuring due diligence} (ensuring they've seen all the options or right results, finding unbiased, credible content, understanding points of view). Participants also found the subsequent process of matching and comparing difficult at times: \textit{Comparing} (comparing products, comparing options of a product), \textit{matching/relevancy} (filtering down options and sifting through results, checking relevance and collecting relevant options, finding a perfect match from options, finding a match that is available to purchase, too much irrelevant content), and \textit{finding the right piece of evidence or item to achieve closure}. Some participants mentioned specific types of sensemaking efforts as difficult, such as \textit{stepping through the process of diagnosis} (e.g. trying various options suggested from the internet, finding relevant instructions in diagnosis tutorials). Finally, participants also found challenges outside of the search process: \textit{Figuring out an initial strategy to approach the problem} (mapping out a plan, figuring out what to search), \textit{executing part of the task outside of searching} (text input, doing math, scheduling, logging in, using a tool, learning a new subject, using a caching service), and \textit{managing external influences on their task} (distraction, boredom, indecision, writing things down, catching up on discord, lack of a specific goal). 

With these in mind, participants also laid out potential solutions that they would invent. These included tools to help digest content focusing on understanding (e.g. simplification, summarization, translation, enhanced audio quality), highlighting key information, general improvements to website layouts and information formats (e.g. less text, shorter text, less videos, more digestible formats), and increased website ease-of-use. Alongside this, participants also mentioned the rapid access of good results, such as 'instant answers', automatic loading of collapsible item descriptions, and generally faster interface loading. Participants also mentioned tools that would help with with processing large amounts of results, specifically focusing on additional filtering/sorting features on websites, increasing result relevancy (e.g. prioritizing good, reputable answers, removing irrelevant, clickbait, or specific results), and tracking options they have already seen. As they dug into these results, participants also requested additional features such as time management, trigger warnings, and random result generators for exploration. In line with exploration, participants also were interested in access to curated results, such as centralized and aggregated information (e.g. a page with everything in one place or single authoritative source), higher level structure (e.g. organized lists, pros and cons, peer experiences, late-stage searches of peers), and pre-rated results (by experts and other similar searchers). This also involved the need for help with high-level search strategies, such as help selecting constraints/queries instead of browsing, lookalike searches, specific guidelines, and example searches. A few participants also lamented the lack of good information (e.g. wishing for more information in reviews or better item descriptions). Lastly, several participants proposed assistants that could help them for various parts of the task, such as consuming (e.g. reading aloud for them), deciding constraints for them, and ranking existing options to constraints.

Reflecting on this list, it is interesting to note that the list of pain points and requested features is long despite a perceived lack of need for tooling. This implies that tools need to bridge this gulf of understanding. People may not understand their process of processing information, as one example, a user examining a nvidia shield controller for purchase may not realize the complex processes (such as comparing, satisfying constraints) that are occurring in their head. For these participants, the process may appear as simple consumption, finding things and reading them. However, consistent with prior work, we find significant complexity in this process and in the decision to continue. People may also  not have salient access to all of this processing, with it happening more implicitly as they feel they engage in their task without significant conscious effort. This suggests a challenge for researchers and designers creating tools to communicate to and educate the intended user of the processes involved.


\section{Limitations}

While we use the term 'sensemaking' as broadly defined by \cite{pirolli_sensemaking_2005} as shorthand to refer to the activities of interest of searching for, collecting, and synthesizing online information, there are many types of sensemaking activities that may not be captured in our findings. Our protocol is targeted to desktop browsing and uses internet browser history as a prompt, which may limit our findings from generalizing across other platforms and applications. In particular, mobile browsing has become a prevalent way for users to process information, and is an important topic for future work. Our method does not log the usage of desktop applications (e.g. Evernote, Slack, Zotero) or assistants (e.g. Cortana, Siri), that other approaches such as desktop screen-recording could capture. Furthermore, within the browser our approach is limited to web history, which means we may not have captured interactions with browser plug-ins and overlays that assist users with knowledge management or specific tasks, although follow-up questions in the survey attempted to mitigate this and our results reporting low usage of external tools suggests that any such assistance was likely limited to a small sample of our participants. Our study asked participants excluded participants who were unable to access their web histories; while this affected a small number of participants, it may result in an under-representation of less-technical users and represents a component that could be improved in future studies. A small number of our participants also could not access their browser history because their primarily browser is configured to automatically delete browser history or was manually cleared during the scope of the study.

Our study was implemented as a two-part modified survey, which as compared to more direct methods such as tool-based or observational studies, can introduce biases in the data collected. We attempted to reduce recall bias by directing participants to reference their search history in answering questions and limited the scope of the study to the previous 72 hours, but participants may have forgotten aspects of their experiences. Web history was never directly recorded but instead volunteered by participants on a task by task basis with options to keep sensitive task details private or omit tasks entirely. In our study, participants used this feature for a very small portion of sensemaking time (<1\%), but it is possible that participants excluded tasks and chose to not report their exclusion to us. Similarly low rates of exclusion in preliminary in-person studies with the protocol gave us some confidence this was not the case. 

Our study recruited US-based crowd participants using the Prolific platform. While a broader sample than university students or knowledge workers, future studies are needed to collect more diverse participants, particularly in non-US countries. The second part of the survey asked users to summarize their task in 4-5 sentences, and we found this to be a useful differentiator between high-effort and low-effort responses. Our study was conducted several months before the general availability of language generation agents such as ChatGPT, so this may not hold for future studies.

Although our methodology captures real users conducting real sensemaking episodes, it is still merely a snapshot over 3 days. We chose to scope our study to 72 hours to reduce cognitive burden and avoid capturing solely weekend data, however it may nonetheless miss long-term user behaviors. A modification of this study could recruit users to take several surveys over a longer period of time similar to a diary study, or re-recruit participants from a global pool opportunistically and generate longitudinal data using a unique identifier. Additionally, as a snapshot, it only represents a frozen and specific period of time. One approach here, hearkening back to the periodic deployments of the WVU WWW survey \cite{pitkow1994results}, could be the establishment of a data collection effort that recruits participants for such surveys on a regular schedule, and could therefore provide data on changing landscapes and emerging tool usage.

\section{Discussion}
\label{sec:discussion}

In our study, we revisited the topic of online research tasks in the wild to better understand the occurrence and composition of everyday sensemaking on the web. In doing so, we empirically quantified the frequency of such tasks and sessions in the daily life of users, finding that users typically engage in several such tasks each day, spending an average of 38 minutes seeking and making sense of information. A small proportion (23\%) of these tasks take up a disproportionate (72\%) amount of time, while about a third (35\%) are concluded in under 5 minutes. Like many previous studies, we found that users are interested in a broad set of domains, but our analysis found that the two categories of Comparison Shopping (29\%) and Researching a hobby/entertainment (24\%) dominated, suggesting that a small set of tool functionality may support a significant proportion of use cases. While many of the forces acting on users in these sessions have been discussed in the literature, our findings provide insight into their relative importance, with users' due diligence and novelty seeking as key drivers - they wanted to do a thorough job (40\%), and felt rewarded finding new results (28\%) that might lead them to something better (25\%). Feeling like they have seen most of the results (saturation) dominated the reasons for stopping, far more than interruption, distraction, or exhaustion.  Overall, our data suggests that sensemaking efforts are rarely explicitly stopped due to completion, but rather left in a state of reasonable satisfaction that may potentially be revisited at a future date. Additional likert scale questions present in the long tasks further confirm this, with only about half of tasks (51.8\%) leaving participants with some amount of closure. These results suggest that the primary reason people do not use support tools is that they don't believe they need them, despite their evident challenges with keeping everything in their heads. We discuss this topic in more detail below. Ultimately, the number and diversity of online research tasks that we found throughout our set of participants suggests that open-ended research tasks have continued to be a common and daily occurrence for users, and continue to frequently involve complex information seeking and sensemaking efforts.

Our results have several implications for the development of new online sensemaking tools. One important factor is motivating how much users actually would need and use such systems in their everyday lives. Although sensemaking and exploratory search has been richly investigated through quantitative (e.g., log analysis) and qualitative (e.g., interview) studies, our study was designed to put together the strengths of each to have a sufficiently long sample (72 hours) of comprehensively user-labeled tasks such that we can estimate frequency and amount of time people spend on what they perceive as open-ended research tasks. 

The answer we find is nuanced in terms of motivating tool development. On the one hand, the amount of time users spend on online sensemaking – broadly defined to include information seeking, collection, organization, and synthesis – appears to be significant at 4.4 hours per week. To provide very rough context, if these results generalized to all internet users (5.3B, https://www.statista.com/statistics/617136/digital-population-worldwide/) this would suggest more than 1 trillion hours of work spent on making sense of online information each year. This figure is consistent with earlier estimates, e.g.,  in \cite{kittur_standing_2014}, which are nearly 10 years old and built on indirect data that is more than 15 years old \cite{kellar_field_2007}. Our work provides fresh, grounded evidence for these estimates, substantiated by users' activities in a newer web landscape - a remarkably consistent ratio despite the substantial changes the web has undergone, and one which suggests strong motivation for the need for tools to help users with such significant time and effort spent. 

On the other hand, our results suggest a paradox in which people don’t think they need such systems. One possible interpretation is that people consistently underestimate the complexity of sensemaking required in an unfamiliar domain until they encounter those complexities, at which point they have sunk significant cost into researching without any tool use, which can make it difficult to motivate starting to use tools. In other words, they feel they are “almost there”, even if they aren’t, similar to findings in cognitive psychology research showing that people are unable to accurately estimate how much time it will take them to solve an insight problem at any point prior to the actual solution \cite{metcalfe1987intuition}. This is supported by more recent lab experiments, showing that people find it difficult to assess how close they are to completing an online search task, across a wide variety of tasks \cite{wu2012grannies} \cite{kelly2015development}. Prior literature regarding the help-seeking behavior in search tasks by Capra et al. \cite{capra2015differences} and orienteering by Teevan et al. \cite{teevan_perfect_2004} further suggest that the "almost there" paradox may affect user's adoption of sensemaking tools through a cascade, triggered by perceptions of low task definition, that causes a self-reinforcing cycle of low tool utility: in the first part of the cascade, the user chooses to forgo tool use believing the task to be too poorly defined for it to be helpful; in the second, the user now has a sufficient understanding of the task (expected solution, required information, associated steps; now nearly a directed search task) to begin seeking assistance, but sees no tool with sufficient context to 'teleport' them to the right answer and abandons the usage of tools a second time to begin orienteering. If true, this would imply that the key to increasing tool usage throughout the research process lies in helping users formulate their thinking with well-defined task descriptions as early as possible (e.g. getting to 'finding a gift for my dog that likes X types of toys using Y sites') and keeping them well-defined as users search (e.g. noting exceptions, noting changes in constraints and preferences, etc). 

Further research is needed to address this seeming paradox. One example might be approaches not requiring effort from the user to capture and instead “looking over their shoulder” (for example, using gaze and cursor position \cite{huang2012user} or collecting and aggregating the options and criteria they encounter \cite{liu_crystalline_2022}) and proactively helping them, especially when it seems they are not finding an immediate point answer. Doing so could also help keep tasks well-defined by stepping in to help users set an explicit boundary to their explorations or agree to a stopping point. Post-hoc tools that could access and leverage the work users have done in previously started research tasks could integrate and reframe results as a continuation of a larger journey instead of making that previous work seem like a sunk cost that isn’t valuable going forward. Such tools could assist users in gradually expanding the scope of their present research by connecting it to previous work, while giving the user sufficient context (e.g. "this exploration is similar to when you browsed OLED TVs last month on eBay and found these two you liked, although last time you primarily looked at prices") to maximize perceptions of task definition, similar to how CiteRead contextualizes research papers \cite{rachatasumrit2022citeread}. Both of these approaches accentuate the continuing contrast between simple information seeking and open-ended exploratory search tasks, with stopping reasons in the latter being more about the perceived confidence in sampling the space than finding an answer. Platforms should consider providing this context about the diversity and distribution of options in the space alongside the current trend of providing users with single answers to satisfy their needs, as knowledge cards and question answering systems are often designed to do. Given the proliferation of business models relying on affiliate marketing and advertising, this may be even more important than it was before as users will rely on the collection of more and more additional content in order to make informed decisions in high 'level of obligation' domains such as healthcare. 

Additionally, it's worthwhile considering how this may change with new approaches such as large language models (e.g. ExploreLLM from Google \cite{ma2023beyond}) that can provide direct answers to everyday open-ended questions. One continuing issue with using LLMs is from the producer side: controlling model hallucinations and enhancing the veracity and verifiability of compiled answers. But a more fundamental question may be on the consumer side: the motivations people reported for continuing their searches had more to do with understanding the space and the distribution of options and tradeoffs in the space than with finding specific point answers. These user needs may still be a mismatch with question answering systems that only focus on providing single answers. However, systems such as the Knowledge Accelerator \cite{hahn_knowledge_2016} show that providing a more comprehensive overview of a space, similar to a meta-analysis of information on any given topic, may address some of these needs and could suggest new ways to interact with such models in the future. For example, replacing the hybrid crowd-AI components of the Knowledge Accelerator with AI-only components may now be possible with techniques such as AI Chaining \cite{wu2022ai}, unlocking lower latency and greater scalability in terms of providing comprehensive overviews of information spaces. However, it is also possible that other ways of providing overviews may be possible with new technologies, such as combining language models with entity or citation graphs (e.g., Synergi \cite{kang2023synergi}), using such models for flexible summarization of elements at scale (e.g. Qlarify \cite{fok2023qlarify} and Scim \cite{fok2023scim}), or other approaches of assembling multiple information processing components such as knowledge retrievers (e.g. Selenite \cite{liu2023selenite}) that have complementary strengths, as the human mind does, rather than relying on a single model to do everything. These more complex outputs would have the additional benefit of enhancing the initial value proposition of tool assistance, which has the potential to improve tool adoption based on previous literature that shows the avoidance of help-seeking behavior in online searching for low-level cognitive tasks such as remembering or recalling facts \cite{capra2015differences}.

\section{Conclusion}

In this paper, we aimed to shed light on sensemaking efforts spent on the modern web through a survey-based approach with aided recall which focuses on segmenting and contextualizing individual exploratory web browsing sessions. This technique enabled a mixed method analysis of everyday sensemaking sessions in the traditional desktop browser setting while preserving user privacy. Using this method, we empirically quantified the frequency of such tasks and sessions in the daily life of 111 participants collected in September, 2022, providing updated evidence for existing estimates in a modern web. Our findings provide insight into the relative importance of different task categories and the forces that act on users in those sessions as accelerators and decelerators of sensemaking efforts. Overall, we found that sensemaking efforts are rarely explicitly stopped due to completion or interruption, but rather left in a state of reasonable satisfaction and later revisited out of obligation to conduct due diligence. Lastly, we found evidence for the lack of tool use in sensemaking sessions despite the evident challenges with keeping everything in their heads, and hypothesize the presence of an 'almost there' effect in which users avoid the use of tools early in the search process until it is too late, and discuss the ensuing implications for future tools and study. As the web environment continues to evolve we hope our snapshot of behavior and survey-based method can be useful to researchers in tracking the evolution of the web environment, whiling providing quantitative and qualitative insights into the contemporary web.

\section{Appendix}
\subsection{Study Protocol}

We attached the study protocol in the following pages of the appendix, formatted as an outline. The survey presentation deviated slightly from the below documented outline due to implementation on the Google Forms platform. 

\begin{figure*}[h]
\centering
\frame{\includegraphics[width=\textwidth]{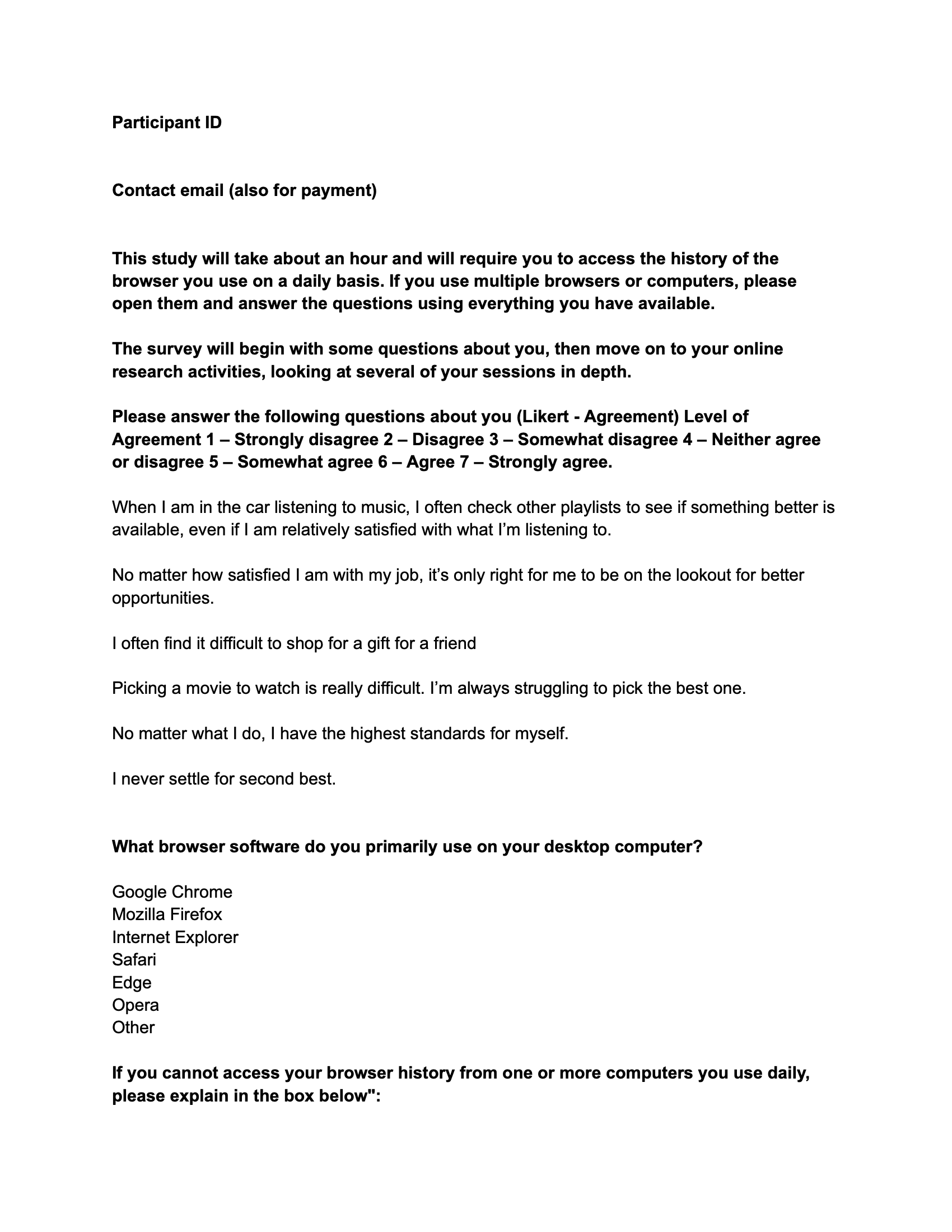}}
\end{figure*}

\begin{figure*}[h]
\centering
\frame{\includegraphics[width=\textwidth]{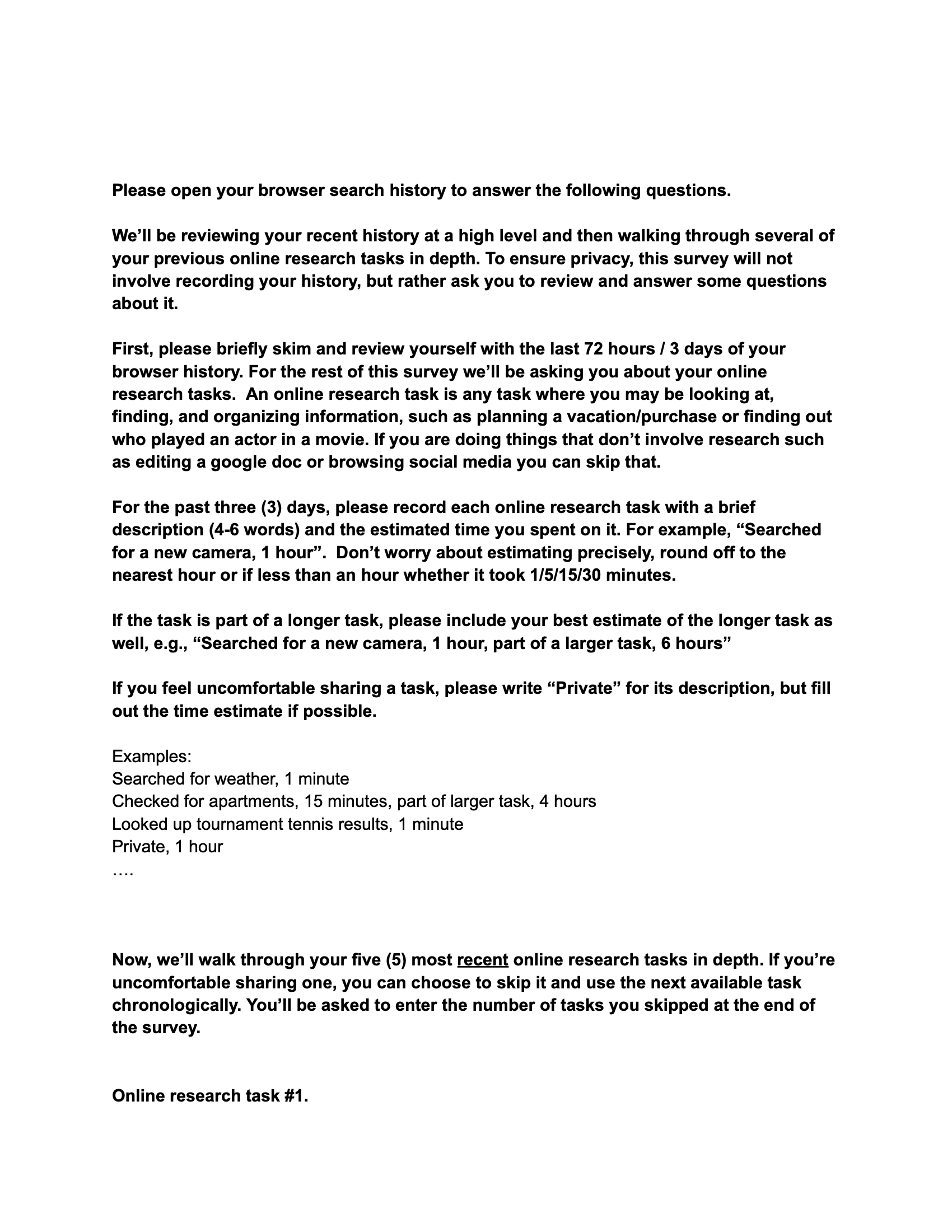}}
\end{figure*}

\begin{figure*}[h]
\centering
\frame{\includegraphics[width=\textwidth]{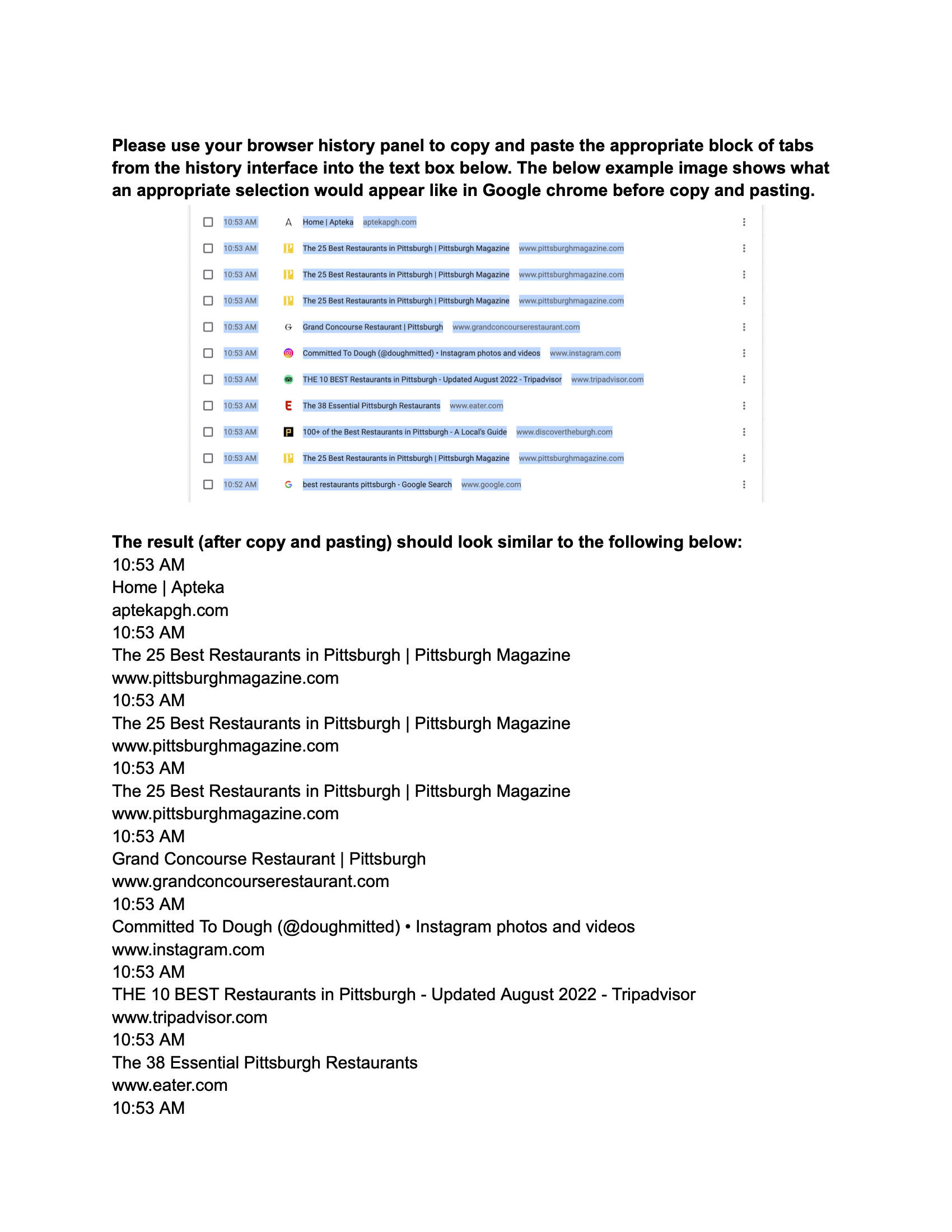}}
\end{figure*}

\begin{figure*}[h]
\centering
\frame{\includegraphics[width=\textwidth]{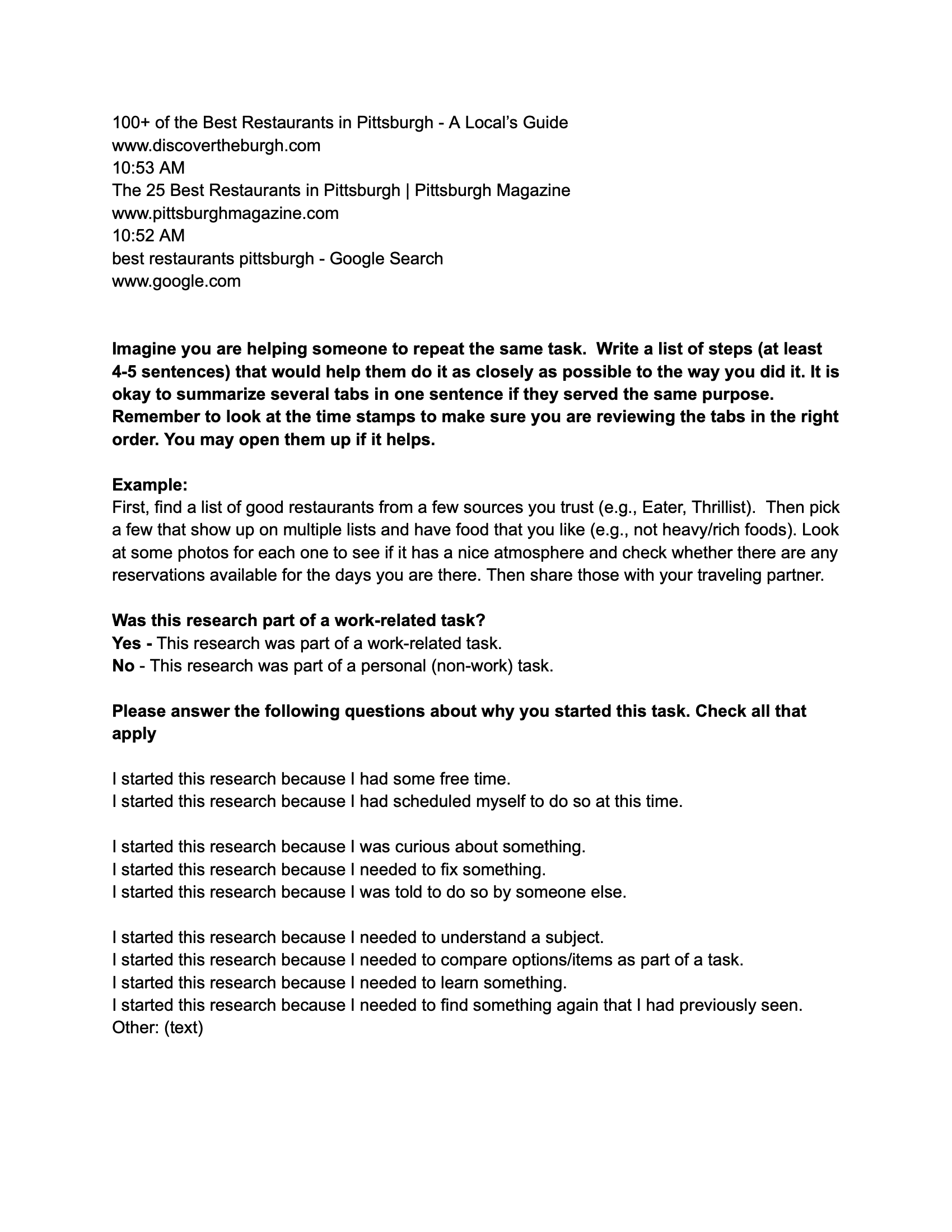}}
\end{figure*}

\begin{figure*}[h]
\centering
\frame{\includegraphics[width=\textwidth]{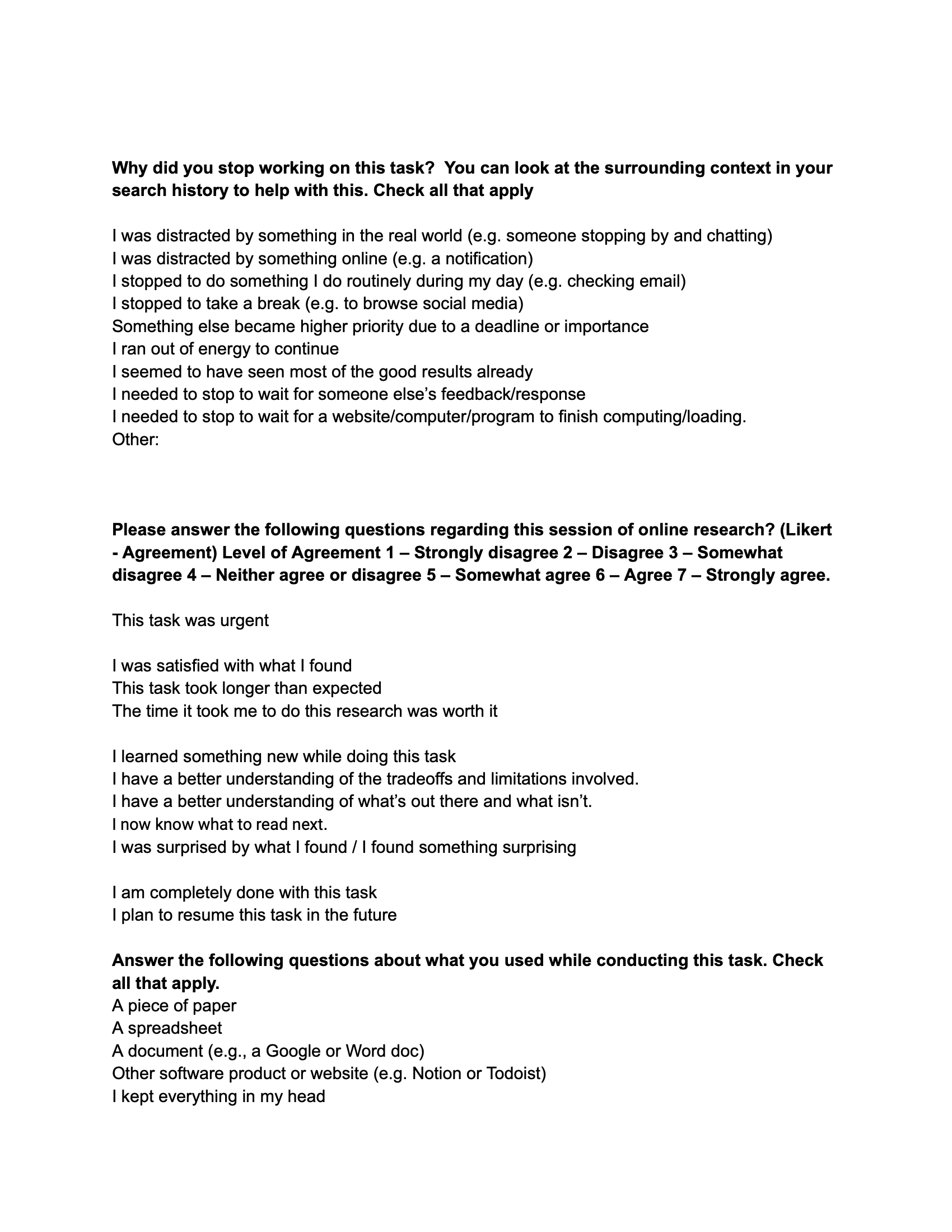}}
\end{figure*}

\begin{figure*}[h]
\centering
\frame{\includegraphics[width=\textwidth]{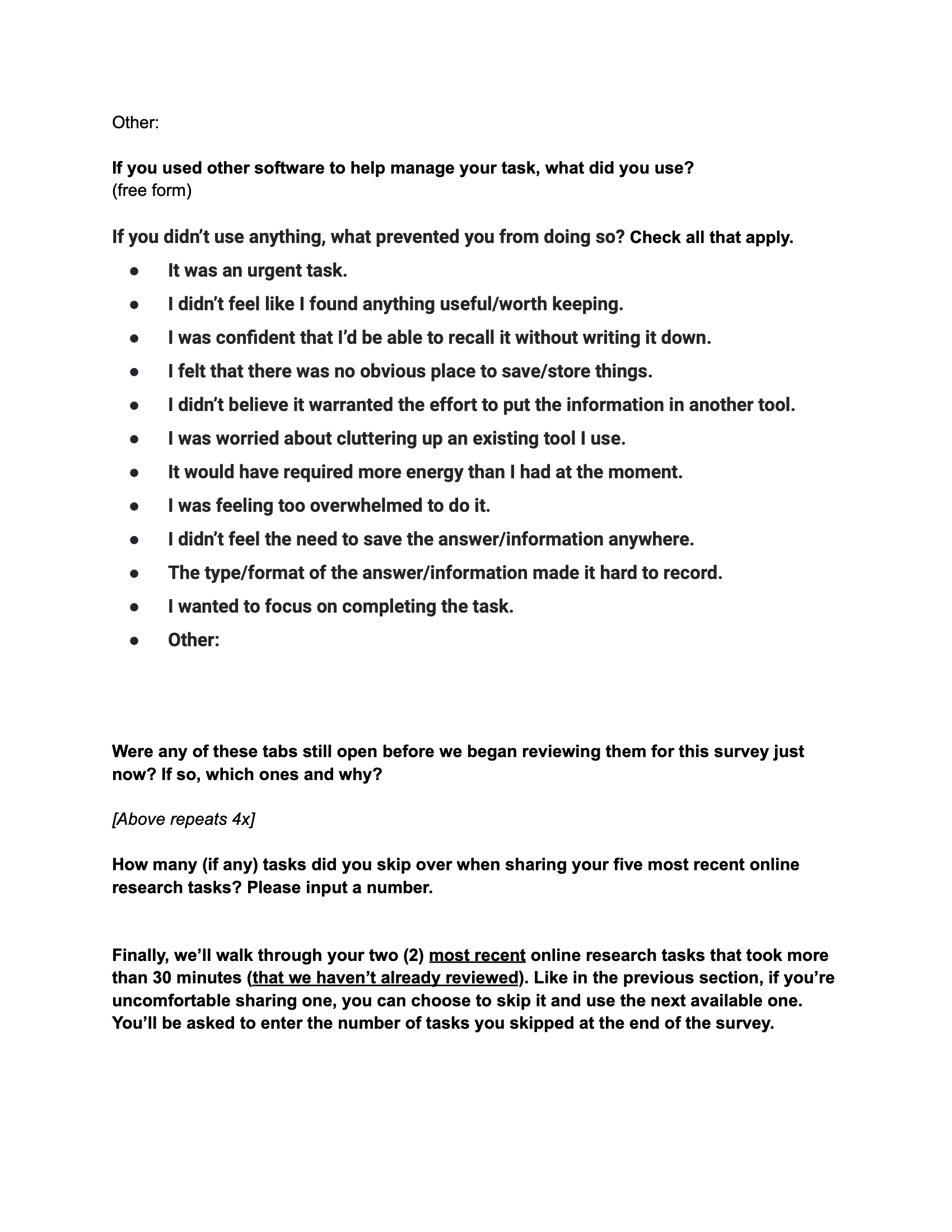}}
\end{figure*}

\begin{figure*}[h]
\centering
\frame{\includegraphics[width=\textwidth]{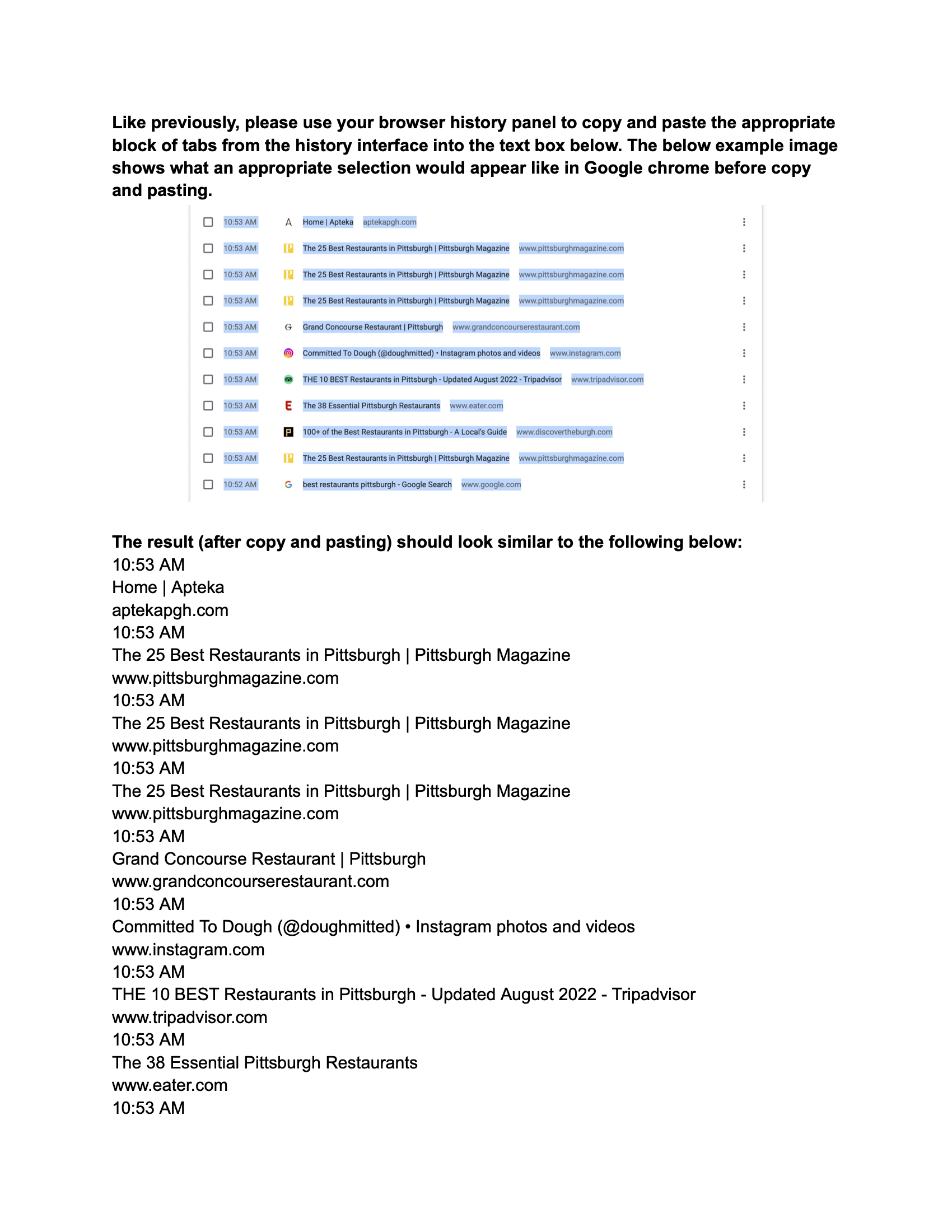}}
\end{figure*}

\begin{figure*}[h]
\centering
\frame{\includegraphics[width=\textwidth]{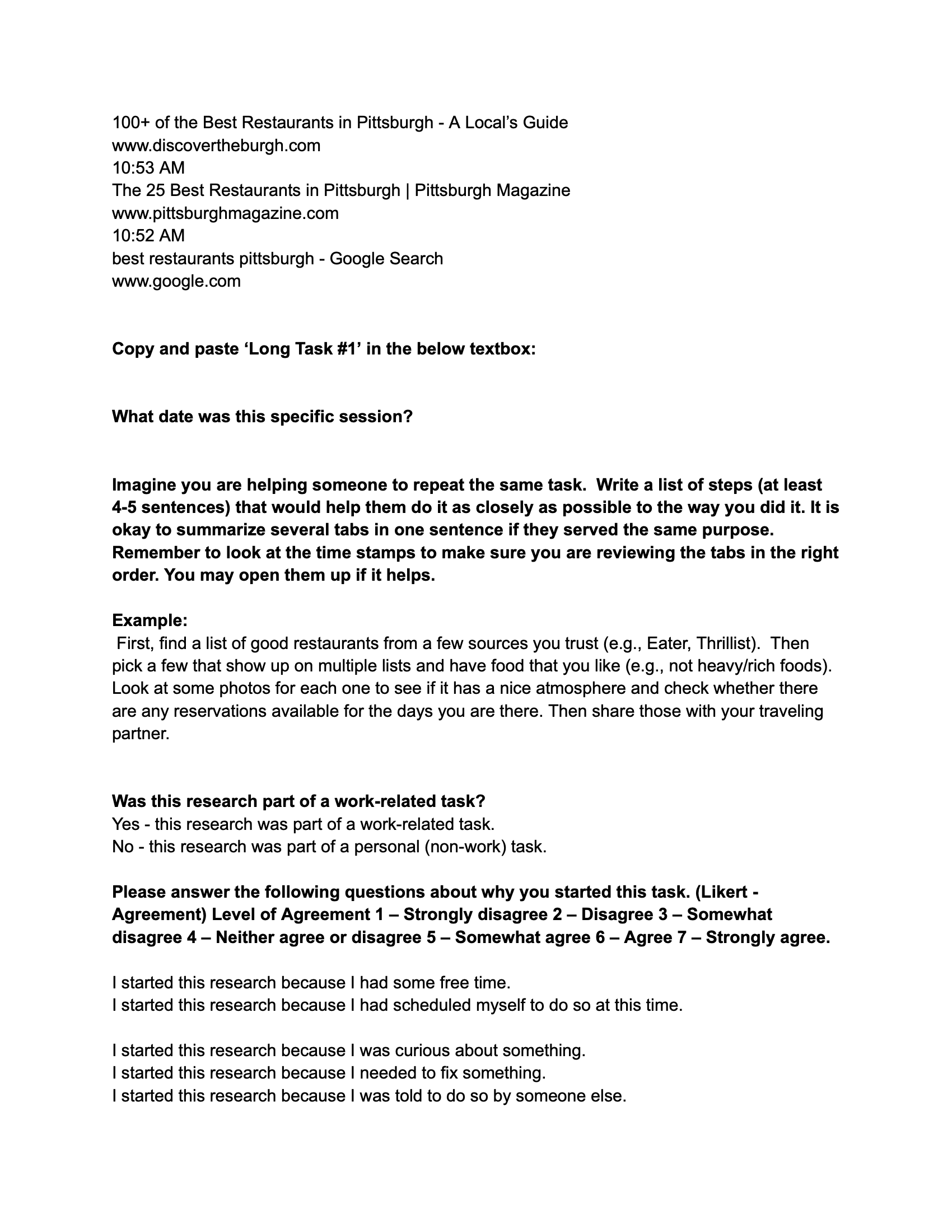}}
\end{figure*}

\begin{figure*}[h]
\centering
\frame{\includegraphics[width=\textwidth]{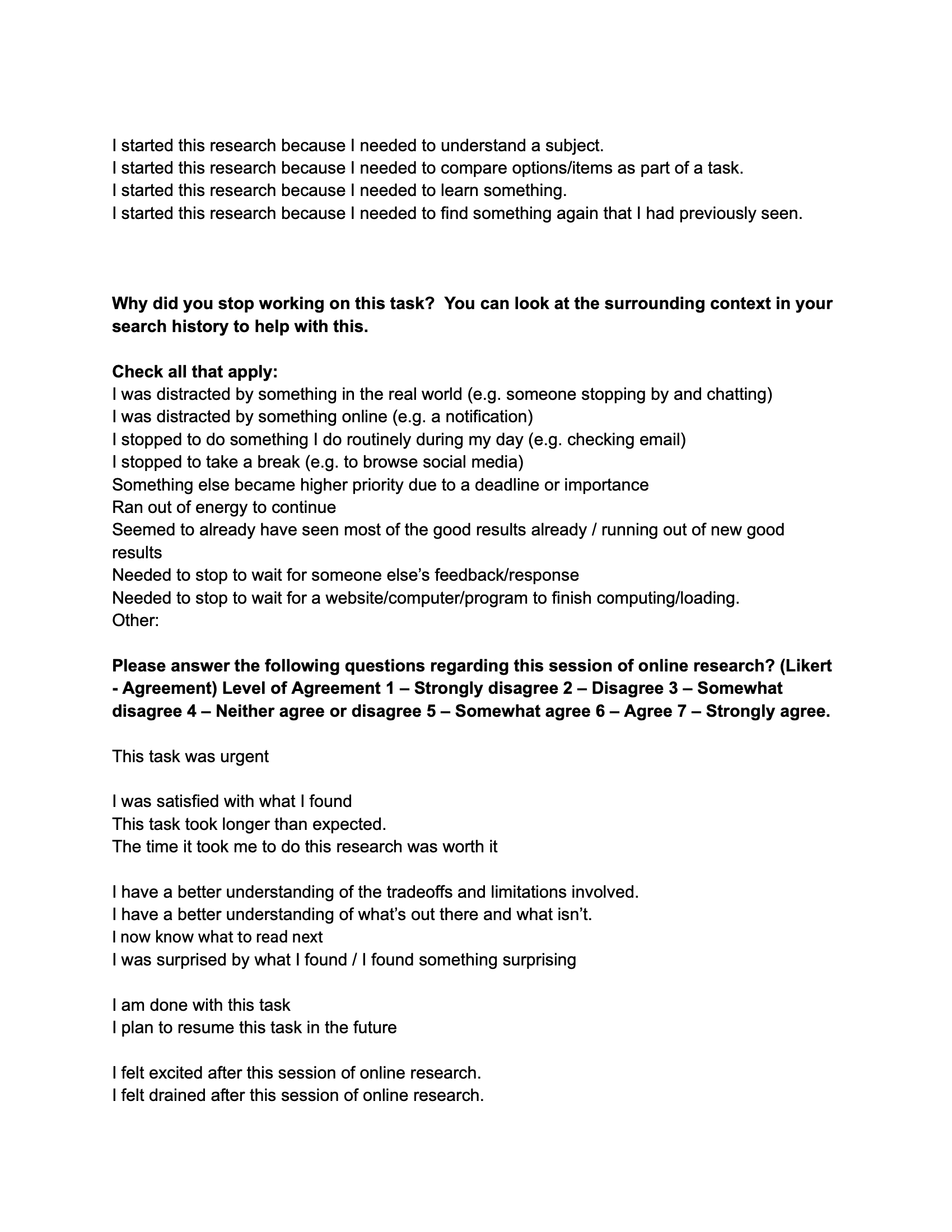}}
\end{figure*}

\begin{figure*}[h]
\centering
\frame{\includegraphics[width=\textwidth]{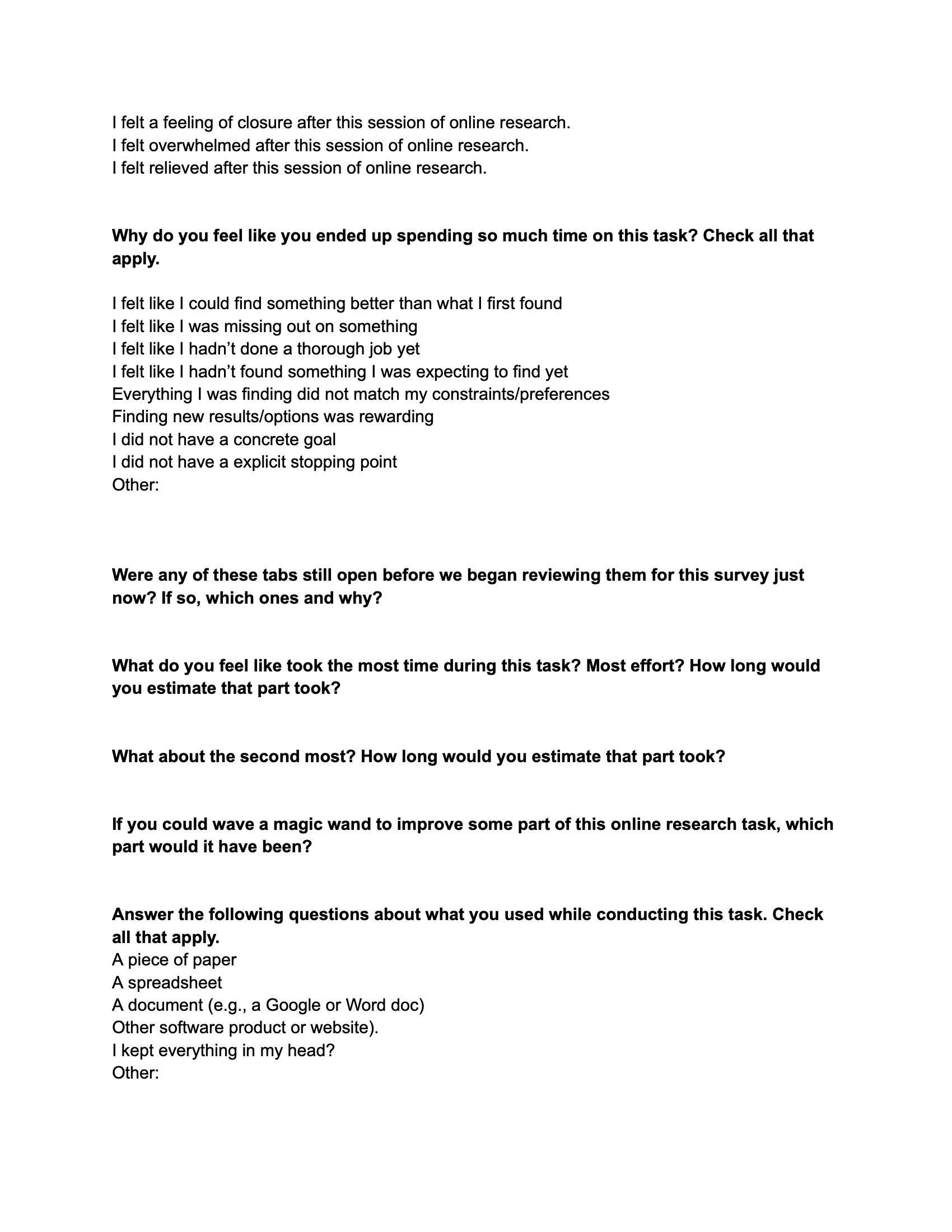}}
\end{figure*}

\begin{figure*}[h]
\centering
\frame{\includegraphics[width=\textwidth]{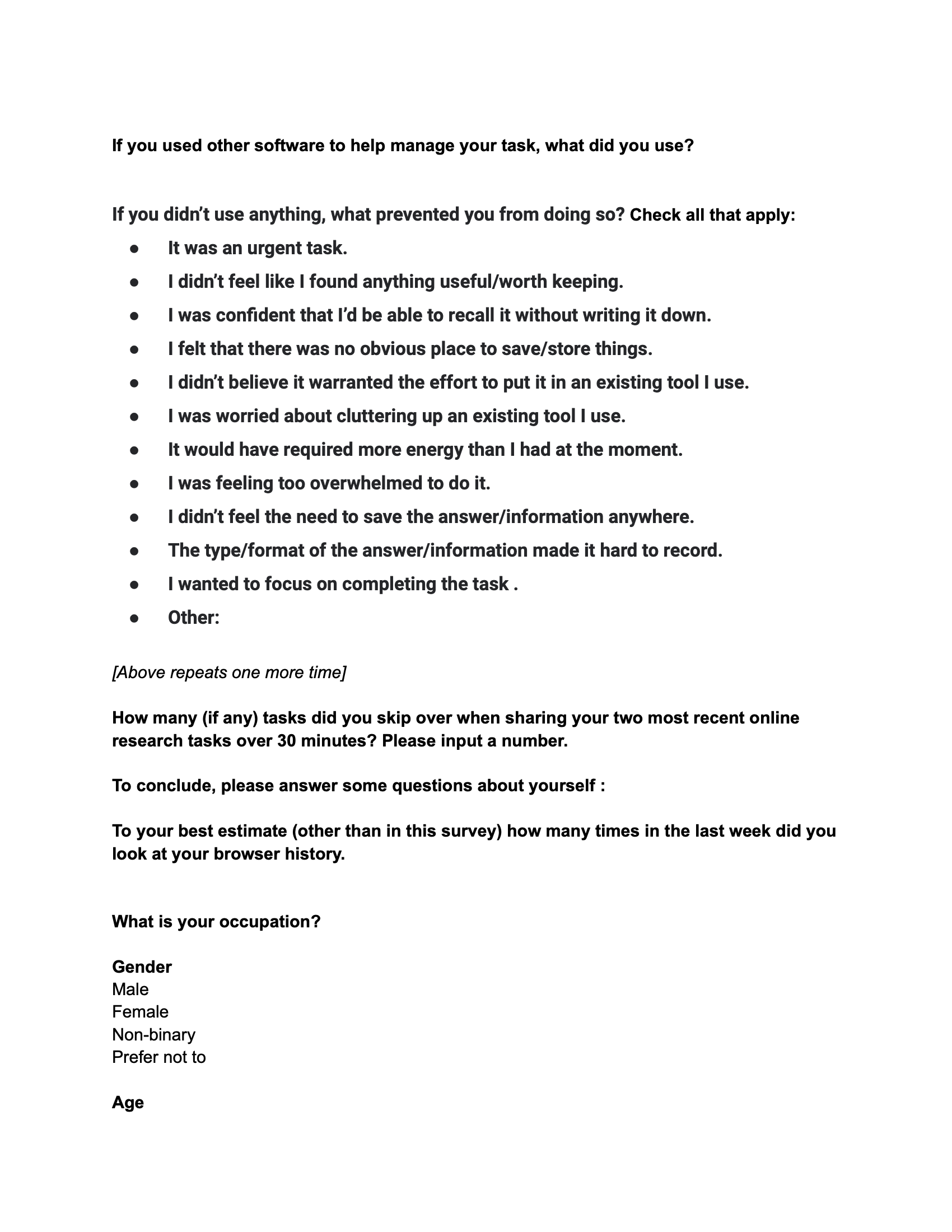}}
\end{figure*}

\begin{figure*}[h]
\centering
\frame{\includegraphics[width=\textwidth]{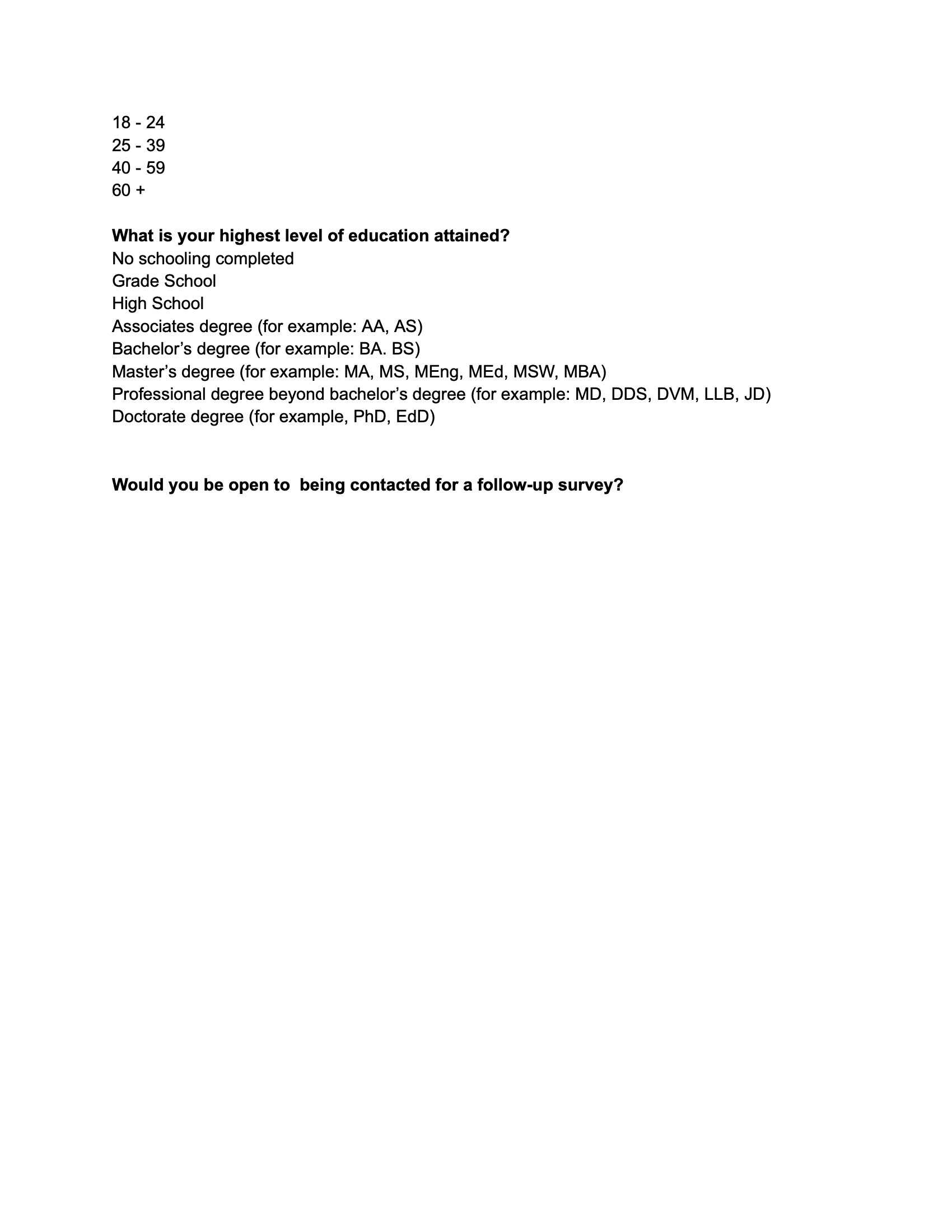}}
\end{figure*}

\clearpage
\begin{acks}
\end{acks}

\bibliographystyle{ACM-Reference-Format}
\bibliography{references,michael-zotero-lib}

\end{document}